\documentclass[prd,superscriptaddress,twocolumn,nofootinbib,showpacs]{revtex4-1}

\usepackage{amsfonts,amssymb,amsmath}
\usepackage{mathrsfs}
\usepackage{color}
\usepackage{graphicx}
\usepackage{dcolumn}

\usepackage{bm}

\newcommand{\C}[1]{{\mathcal #1}}

\newcommand{\BS}[1]{{\boldsymbol #1}}
\newcommand{\BF}[1]{{\mathbf #1}}
\newcommand{\SC}[1]{{\mathscr #1}}

\newcommand{\beq}{\begin{equation}}
\newcommand{\eeq}{\end{equation}}
\newcommand{\bea}{\begin{eqnarray}}
\newcommand{\eea}{\end{eqnarray}}
\newcommand{\nn}{\nonumber}

\newcommand{\Tr}{\mathop{\rm Tr}}

\newcommand{\half}{\frac 12}
\newcommand{\third}{\frac 13}

\newcommand{\sixth}{\frac 16}

\newcommand{\Slash}[1]{{\ooalign{\hfil#1\hfil\crcr\raise.167ex\hbox{/}}}}

\begin{document}

\title{Multifield dynamics of supersymmetric Higgs inflation in $SU(5)$ GUT}

\author{Shinsuke Kawai}
\email{kawai@skku.edu}
\affiliation{Department of Physics, Sungkyunkwan University,
Suwon 440-746, Republic of Korea}
\author{Jinsu Kim}
\email{kimjinsu@skku.edu}
\affiliation{Department of Physics, Sungkyunkwan University,
Suwon 440-746, Republic of Korea}

\date{\today}

\begin{abstract}
We study the Higgs inflation model realized in the supersymmetric $SU(5)$ grand
unified theory (GUT), focusing on its multifield dynamics and prediction of
cosmological observables.
The requirement for GUT symmetry breaking during inflation imposes tight constraints on the model parameters.
We find, nevertheless, with an appropriately chosen noncanonical K\"ahler potential the model is in excellent agreement with the present cosmological observation.
The effects from multifield dynamics is found to be minor and thus,
unlike other similar supersymmetric implementation of nonminimally coupled Higgs inflation, the prediction of this model is robust against multifield ambiguities.
\end{abstract}

\maketitle

\section{Introduction}\label{sec:intro}
%
The quest for a concrete particle theory realization of cosmological inflation
continues to be a major theoretical challenge.
Current experiments put stringent bounds on the amplitude of the tensor mode
primordial fluctuations.
To quote the results of the BICEP2/Keck Array and Planck collaborations
\cite{Ade:2015xua},
the primordial tensor-to-scalar fluctuation ratio is constrained to be
\begin{align}\label{eqn:rBKP}
r_{0.002}< 0.09
\end{align}
({\em Planck} TT+lowP+lensing+ext+BKP) at 95\% confidence.
The simple chaotic inflation models with a quadratic or quartic inflaton potential are disfavored by the observation.
Instead, the $R^2$-inflation model \cite{Nariai:1971sv,Starobinsky:1980te} and the nonminimally coupled Higgs inflation model
\cite{CervantesCota:1995tz,Bezrukov:2007ep},
among others, have (re)surfaced as viable accounts of the early Universe.
In particular, the Bezrukov-Shaposhnikov scenario \cite{Bezrukov:2007ep} of nonminimally coupled Higgs inflation is an attractive proposal.
This scenario is economical as there is no need to introduce a new field of unknown origin; the Higgs field that already exists in the Standard Model (SM) is responsible for inflation.
The model also has strong predictive power as the physics at the inflationary scale is related to that of the collider scale through the renormalization group flow \cite{DeSimone:2008ei}.
There is a controversy on the unitarity problem associated with the large nonminimal coupling required in this type of scenario \cite{Burgess:2009ea,Burgess:2010zq,Barbon:2009ya}.
This danger may be avoided, e.g. by considering the cutoff scale as field-dependent \cite{Bezrukov:2010jz}.

%
The Bezrukov-Shaposhnikov scenario of Higgs inflation assumes that the SM is valid all the way up to the energy scale of inflation.
However, it is widely believed that the grand unification takes place at the energy scale of
$M_{\rm GUT}\sim 10^{16}$ GeV and there the physics is supposed to be described by grand unified theory (GUT).
The tensor-to-scalar ratio of $r\approx 0.05$, for example\footnote{
The BICEP2/Keck Array/Planck joint analysis
\cite{Ade:2015tva} gives $r=0.048^{+0.035}_{-0.032}$ at 68\% confidence.
}, implies that the Hubble parameter during inflation can be as large as $H\sim 10^{14}$ GeV;
this is closer to the GUT scale than to the electroweak scale, and thus inflation may be more appropriately discussed in the framework of GUT than in the SM.
In view of the elegant gauge coupling unification in the presence of supersymmetry, a natural
beyond-the-SM extension of the Bezrukov-Shaposhnikov scenario 
would be in supersymmetric GUT.

Implementation of the nonminimally coupled inflationary scenario in supersymmetric GUT has been discussed e.g. in the supersymmetric $SU(5)$ model\footnote{
Nonminimally coupled $SU(5)$ GUT Higgs inflation without supersymmetry is discussed in \cite{CervantesCota:1994zf}.
} \cite{Arai:2011nq} and the supersymmetric Pati-Salam model \cite{Pallis:2011gr}.
By construction, these nonminimally coupled models employ a noncanonical K\"{a}hler potential to circumvent the supergravity $\eta$ problem and at the same time to suppress the tensor mode fluctuations to be compatible with the observation \eqref{eqn:rBKP}.
This type of model involves multiple scalars and in principle, the prediction for cosmological observables depends on the trajectory of the inflaton in the multidimensional field space.
As pointed out in \cite{Einhorn:2009bh,Ferrara:2010yw,Einhorn:2012ih}, there is a danger of tachyonic instabilities in undesired directions of the field space, but the instabilities may be removed and the inflaton trajectory can be controlled by further noncanonical terms in the K\"{a}hler potential \cite{Ferrara:2010in}
(see also \cite{Arai:2011aa,Arai:2012em,Kawai:2014doa,Arai:2013vaa}).
Thus the trajectory of the inflaton is in general sensitive to the noncanonical terms, and naturally the prediction for the cosmological observables depends on the K\"{a}hler potential.
Conversely, the current observational constraints may be used to restrict the parameter space of the K\"{a}hler potential \cite{Kawai:2014gqa}.

In this paper we study the multifield dynamics of supersymmetric GUT-embedded nonminimally coupled Higgs inflation.
Our main focus is on the model based on the {\em minimal} supersymmetric $SU(5)$ GUT;
this is the simplest Higgs inflation model in supersymmetric GUT that involves symmetry breaking of a GUT group, and hence it serves as a prototype of GUT-based Higgs inflation models.
The scenario was discussed in \cite{Arai:2011nq} using a crude single-field approximation.
The purpose of the present paper is to analyze it appropriately as a multifield inflationary model and reinvestigate its predictions.
This model differs from the similar supersymmetric Higgs inflation models implemented in the next-to-minimal supersymmetric Standard Model (NMSSM) \cite{Ferrara:2010yw} or the supersymmetric seesaw model \cite{Arai:2011aa,Arai:2012em,Kawai:2014doa}, by the form of the K\"{a}hler metric, which is essential for the phenomenological consistency of the GUT model as the GUT symmetry needs to be broken.
We investigate cosmological consequences of this feature.

In the next section we start by reviewing the supersymmetric Higgs inflation model based on the $SU(5)$ GUT \cite{Arai:2011nq}.
For the sake of concreteness we focus on the two-field case arising from the supersymmetric minimal $SU(5)$ GUT, and describe its construction in detail.
We analyze the model in Sec. \ref{sec:numerics} and present the numerical results.
In Sec. \ref{sec:concl} we conclude, with brief discussions on our results.
The technicalities of the multifield inflationary dynamics are summarized in Appendix. \ref{sec:dynamics}.
%

\section{Higgs inflation in $SU(5)$ GUT}\label{sec:model}
%
We recall that the Georgi-Glashow $SU(5)$ GUT 
\cite{Georgi:1974sy} consists of
the gauge field in ${\BS 2\BS 4}$,
the GUT Higgs field in ${\BS 2\BS 4}$,
the SM Higgs field in ${\BS 5}$,
$N_F$ (the flavor multiplicity) fermion fields in ${\BS 1\BS 0}$ and
$N_F$ fermion fields in $\overline{\BS 5}$,
in the representations of $SU(5)_{\rm GUT}$.
The gauge and the SM Higgs fields decompose into the representations of the SM gauge group
$SU(3)_c\times SU(2)_L\times U(1)_Y$ as
\begin{align}
{\BS 2\BS 4}=&\underbrace{({\BS 8}, {\BS 1}, 0)}_{\text{gluon}}
+\underbrace{({\BS 1}, {\BS 3}, 0)}_{A^a_\mu}
+\underbrace{({\BS 1}, {\BS 1}, 0)}_{B_\mu}
+\underbrace{({\BS 3}, {\BS 2}, \frac 56)}_{X^\alpha_\mu,Y^\alpha_\mu}
+\underbrace{(\overline{\BS 3}, {\BS 2}, -\frac 56)}_{\overline X^\alpha_\mu,\overline Y^\alpha_\mu},\nn\\
\label{eqn:5}
{\BS 5}=&\underbrace{({\BS 3}, {\BS 1}, -\third)}_{\Phi_T}
+\underbrace{({\BS 1}, {\BS 2}, \half)}_{\Phi_D}.
\end{align}
Here, $\mu=0,1,2,3$ is the spacetime index, $a=1,2,3$ is the $SU(2)_L$ index, $\alpha=1,2,3$ is the color index, $\Phi_T$ is the colored (triplet) Higgs and $\Phi_D$ is the SM (doublet) Higgs.
Since the color symmetry is unbroken in the SM vacuum, $\langle\Phi_T\rangle=0$.
The SM Higgs vacuum expectation value (VEV) is $\langle\Phi_D\rangle=246$ GeV.
The GUT Higgs field breaks the GUT symmetry down to the SM symmetry by giving GUT scale masses to the $X$, $Y$, $\overline X$, $\overline Y$ fields.
In the representations of $SU(3)_c\times SU(2)_L$, the fermion fields are
\begin{align}
{\BS 1\BS 0}=&\underbrace{({\BS 1},{\BS 1})}_{e^c}
+\underbrace{(\overline{\BS 3},{\BS 1})}_{u^c}
+\underbrace{({\BS 3},{\BS 2})}_{Q=(u, d)_L},\crcr
\overline{\BS 5}=&\underbrace{(\overline{\BS 3},{\BS 1})}_{d^c}
+\underbrace{({\BS 1},{\BS 2})}_{L=(e,\nu_{e})_L}.
\end{align}
In the supersymmetric $SU(5)$ GUT, there are two Higgs doublets
$H_1\equiv H_d$ and $H_2\equiv H_u\supset \Phi_D$.
The field contents of the minimal $SU(5)$ model are one vector supermultiplet in ${\BS 2\BS 4}$ and 5 kinds of chiral supermultiplets:
\begin{itemize}
  \item{$\Sigma$} in ${\BS 2\BS 4}$ (GUT Higgs),
  \item{$H$} in ${\BS 5}$ (including $H_u$),
  \item{$\overline H$} in $\overline{\BF 5}$ (including $H_d$),
  \item{$N_F$ families of $\chi_{ij}$} in ${\BF 1\BF 0}$ (including $Q$, $u^c$ and $e^c$),
  \item{$N_F$ families of $\eta^i$} in $\overline{\BF 5}$ (including $L$ and $d^c$).
\end{itemize}
The inflationary model we discuss involves only $\Sigma$, $H$, and $\overline H$;
we will neglect the vector multiplet ${\BS 2\BS 4}$ and the chiral multiplets
$\chi_{ij}$ (${\BF 1\BF 0}$) and $\eta^i$ ($\overline{\BF 5}$) below.

\subsection{Superpotential for $SU(5)$ GUT Higgs inflation}\label{sec:superpotential}

We consider the GUT superpotential given by
\begin{align}\label{eqn:W}
W=\overline H (\mu+\rho\Sigma) H+\frac{m}{2}\Tr (\Sigma^2)+\frac{\lambda}{3}\Tr (\Sigma^3),
\end{align}
where $\mu$, $\rho$, $m$, $\lambda$ are real constant parameters.
The scalar component of $\Sigma$ is a traceless $5\times 5$ matrix $\Sigma^i{}_j$.
For the (almost) canonical K\"{a}hler metric the potential constructed from the second and the
third terms of \eqref{eqn:W} has three distinct vacua:
\begin{align}
\Sigma^i{}_j=&~ 0,\nn\\
\Sigma^i{}_j=&~ \frac{m}{3\lambda}{\rm diag}(1,1,1,1,-4),\nn\\
\Sigma^i{}_j=&~ \frac{m}{\lambda}{\rm diag}(2,2,2,-3,-3).
\end{align}
The first vacuum corresponds to the unbroken $SU(5)$,
the second corresponds to the spontaneous symmetry breaking
$SU(5)\to SU(4)\times U(1)$, and the third one to
$SU(5)\to SU(3)\times SU(2)\times U(1)$.
Obviously, for the SM particle physics to be realized after inflation we need
the last configuration of the $\Sigma$ field.
We use a singlet chiral superfield $S$ to write
\begin{align}
\Sigma=\sqrt{\frac{2}{15}}\, S\, {\rm diag}\left(
1, 1, 1, -\frac 32, -\frac 32\right).
\end{align}
It can be easily verified that $\Tr(\Sigma^2)=S^2$, $\Tr(\Sigma^3)=-\frac{1}{\sqrt{30}}S^3$, $\Tr(\Sigma^\dag\Sigma)=|S|^2$, etc.
Writing the ${\BS 5}$ and $\overline{\BS 5}$ Higgs multiplets as
\begin{align}
H=\left(\begin{array}{c}H_c,\cr H_u\end{array}\right),
\quad
\overline H=\left(\begin{array}{c}\overline H_c,\cr H_d\end{array}\right),
\end{align}
in which the bosonic parts of $H_c$ and $H_u$ are $\Phi_T$ and $\Phi_D$ in \eqref{eqn:5},
the superpotential becomes
\begin{align}\label{eqn:W1}
W=& \left(\mu+\sqrt{\frac{2}{15}}\rho S\right)\overline H_c H_c
+\left(\mu-\sqrt{\frac{3}{10}}\rho S\right) H_u H_d\nn\\
&+\half m S^2 -\frac{\lambda}{3\sqrt{30}} S^3.
\end{align}
In the GUT scale the SM Higgs field is almost massless, requiring
\begin{align}\label{eqn:mu}
\mu-\sqrt{\frac{3}{10}}\, \rho\, \langle S\rangle=0.
\end{align}
The color Higgs fields must have vanishing VEV, $\langle\overline H_c\rangle=\langle H_c\rangle=0$ since the color
symmetry is unbroken throughout the history.
They are expected to have GUT scale masses and hence
\begin{align}\label{eqn:cHiggsCond}
\mu+\sqrt{\frac{2}{15}}\, \rho\, \langle S\rangle\simeq M_{\rm GUT}=2\times 10^{16} \text{ GeV}.
\end{align}
The conditions \eqref{eqn:mu} and \eqref{eqn:cHiggsCond} lead to
$\mu\simeq\rho\langle S\rangle\simeq M_{\rm GUT}$.

Near the SM vacuum ($H_c=\overline H_c=0$, $H_u, H_d\ll M_{\rm GUT}$) we further impose a stationarity condition 
\begin{align}
\frac{\delta W}{\delta S}= S\left(m-\frac{\lambda}{\sqrt{30}}S\right)=0.
\end{align}
Since $\langle S\rangle\neq 0$, we must have
\begin{align}\label{eqn:m}
m-\frac{\lambda}{\sqrt{30}}\langle S\rangle=0.
\end{align}
Denoting $\widetilde v\equiv\langle S\rangle$, the conditions \eqref{eqn:mu} and \eqref{eqn:m} give
\begin{align}
\mu= \sqrt{\frac{3}{10}}\, \rho\, \widetilde v,
\quad
m=\frac{\lambda\, \widetilde v}{\sqrt{30}}.
\end{align}
In the Higgs doublets the charged components can be consistently set to zero,
\begin{align}
H_u=\left(
\begin{array}{c} 0\cr H^0_u\end{array}\right),
\quad
H_d=\left(
\begin{array}{c} H^0_d\cr 0\end{array}\right).
\end{align}
Recalling that the contraction of the $SU(2)$ doublets uses the $SU(2)$ invariant
$i\sigma_2=${\scriptsize$ \left(\begin{array}{cc} 0 & 1\\ -1 &0\end{array}\right)$},
we have
$H_u H_d=H_u^T i\sigma_2 H_d=-H_u^0 H_d^0$.
With $H_c=\overline H_c=0$, the superpotential \eqref{eqn:W1} now reads
\begin{align}
W=\sqrt{\frac{3}{10}}\rho (S-\widetilde v) H_u^0 H_d^0 + \half m S^2-\frac{m}{3\widetilde v}S^3.
\end{align}
This is the superpotential we shall use for the inflationary model.

\subsection{The cubic K\"{a}hler model}\label{sec:cubic}
For successful cosmological inflation the inflaton potential needs to satisfy at least the following three conditions:
(i) sufficiently flat so that slow roll takes place;
(ii) exhibits no tachyonic instabilities in the directions orthogonal to the desired trajectory;
(iii) the inflaton trajectory settles at the SM vacuum after the slow roll.
The difficulty to achieve (i) within supergravity is known as {\em the $\eta$ problem}.
One way to circumvent the $\eta$ problem is to use a noncanonical K\"{a}hler potential\footnote{
The $\eta$ problem states that assuming the canonical K\"{a}hler potential, a generic superpotential and F-term supersymmetry breaking, the slow roll parameter $\eta$ can never be $\ll {\C O}(1)$.
Therefore it may be avoided also by considering D-term supersymmetry breaking or using a specially engineered form of the superpotential.
See \cite{Yamaguchi:2011kg} for a review.
}, and here we use, following \cite{Arai:2011nq}, the K\"{a}hler potential
$K=-3\ln\Phi$, where
\begin{align}\label{eqn:cubicPhi1}
\Phi &=1-\third\left(
\Tr\Sigma^\dag\Sigma + |H|^2+|\overline H|^2+\cdots\right)\crcr
&\quad
-\frac{\gamma}{2}\left(
\overline H H+H^\dag \overline H^\dag\right)
+\frac{\widetilde\omega}{3}\left(
\Tr\Sigma^\dag\Sigma^2+\Tr\Sigma^{\dag 2}\Sigma\right)\crcr
&\quad
+\frac{\zeta}{3}\left(
\Tr\Sigma^\dag\Sigma\right)^2.
\end{align}
The reduced Planck mass $M_{\rm P}=2.44\times 10^{18}$ GeV is set to unity.
As we shall see, the conditions (ii) and (iii) above will also be fulfilled if the real parameters
$\widetilde\omega$ and $\zeta$ are chosen appropriately.
The ellipsis in the first line of \eqref{eqn:cubicPhi1} represents the canonical terms for the superfields other than
$\Sigma={\BS 2\BS 4}$, $H={\BS 5}$, $\overline H=\overline{\BS 5}$.
Canonical here is in the sense of the superconformal framework
\cite{Kaku:1978nz,Siegel:1978mj,Cremmer:1982en,Ferrara:1983dh,Kugo:1982mr,Kugo:1982cu,Kugo:1983mv,Freedman:2012zz}, in which the K\"{a}hler metric constructed from the superconformal K\"{a}hler potential ${\C K}\equiv -3\Phi$ becomes trivial.
The terms in the second and the third lines are noncanonical.
The term proportional to $\gamma$ renders the potential to be flat, in a manner
analogous to the nonminimal coupling in the SM Higgs inflation model.
The quartic term (proportional to $\zeta$) controls the tachyonic instability, and the cubic terms (proportional to $\widetilde\omega$) control the symmetry of the potential so that the SM vacuum can be reached after the slow roll.
The function $\Phi$ can be written using the component fields as
\begin{align}\label{eqn:cubicPhi2}
\Phi &= 1-\third\left(
|S|^2+|H_u^0|^2+|H_d^0|^2\right)
+\frac{\gamma}{2}\left(H_u^0H_d^0+\text{c.c.}\right)\crcr
&\quad
-\frac{\widetilde\omega}{3\sqrt{30}}\left(\overline S S^2+\overline S^2 S\right)
+\frac{\zeta}{3}|S|^4.
\end{align}

\subsubsection{Jordan frame}\label{sec:cubicJordan}

To proceed, we consider the D-flat direction along $H_u$-$H_d$ and parametrize it by a superfield $\varphi$ as
\begin{align}
H_u^0=\frac{1}{\sqrt 2}\varphi,
\quad
H_d^0=\frac{1}{\sqrt 2}\varphi.
\end{align}
Further, it is convenient to rescale the scalar components of $S$ and $\varphi$ as\footnote{
The normalization of $s$, $\omega$, $v$ differs from \cite{Arai:2011nq} by a factor of $\sqrt 2$:
$s_{\rm there} = s_{\rm here}/\sqrt{2}$, $\omega_{\rm there} = \omega_{\rm here}/\sqrt{2}$,
$v_{\rm there}=\widetilde v_{\rm here}= v_{\rm here}/\sqrt{2}$.
The superconformal K\"{a}hler potential ${\C K}$ is written in \cite{Arai:2011nq} as $K$.
Note that $K=-3\ln (-{\C K}/3)$ in this paper.
}

\begin{align}
S=\frac{1}{\sqrt 2}s,
\quad
\varphi=\frac{1}{\sqrt 2}h.
\end{align}
With this normalization, the scalar-gravity part of the Lagrangian density takes the following form
\begin{align}\label{eqn:LJ}
{\SC L}_{\rm J}=\sqrt{-g_{\rm J}}\left[
\frac{\Phi{\SC R}_{\rm J}}{2}
-\half g_{\rm J}^{\mu\nu}\partial_\mu h\partial_\nu h
-\frac{\kappa}{2} g_{\rm J}^{\mu\nu}\partial_\mu s\partial_\nu s -V_{\rm J}\right],
\end{align}
where $g_{\rm J}^{\mu\nu}$ is the inverse of the Jordan frame spacetime metric $g^{\rm J}_{\mu\nu}$, ${\SC R}_J$ is the scalar curvature in the Jordan frame 
and
\begin{align}
\Phi=&1-\sixth s^2+\sixth\omega s^3+\frac{\zeta}{12} s^4+\left(\frac{\gamma}{4}-\sixth\right) h^2,\\
\kappa\equiv &1-2\omega s-2\zeta s^2,\label{eqn:kappacubic}\\
\omega\equiv &-\frac{\widetilde\omega}{\sqrt{15}}.
\end{align}
The scalar potential $V_{\rm J}$ in the Jordan frame is the F-term potential
\begin{align}
V_{\rm F}
=\frac{W_i ({\C K}\overline W_{\overline\jmath}-3{\C K}_{\overline\jmath}\overline W)
-3W({\C K}_i\overline W_{\overline\jmath}-3 {\C K}_{i\overline\jmath}\overline W)
}{{\C K}{\C K}_{i\overline\jmath}-{\C K}_i{\C K}_{\overline\jmath}},
\end{align}
where the subscripts $i$ and $\overline\jmath$ denote differentiation with respect to the chiral and anitchiral superfields.
In terms of the component fields its explicit form is
\begin{widetext}
\begin{align}\label{eqn:VJcubic}
V_{\rm J}
=
\frac{3}{40}\left\{\rho^2(s-v)^2h^2
+\frac{1}{\kappa}\left[\frac{\rho h^2}{2}-\frac{\lambda}{3} s(s-v)\right]^2\right\}
-\frac{\left\{\frac{2\zeta s+\omega}{\kappa}\!
\left[\frac{\rho}{2}h^2
\! -\!\frac{\lambda}{3}s(s-v)\right]\! s^2\!
+\rho v h^2
-\frac{\lambda}{3}v s^2
-3\gamma\rho h^2 (s-v)\right\}^2}
{80\left[4+\gamma(\frac 32 \gamma-1) h^2+\frac{2\zeta+\omega^2}{6\kappa} s^4
\right]},
\end{align}
\end{widetext}
where we have introduced $v\equiv {\sqrt 2} \widetilde v$.

\subsubsection{Einstein frame}\label{sec:cubicEinstein}
To discuss cosmology it is convenient to bring the Lagrangian \eqref{eqn:LJ} to the Einstein frame
in which the fields are minimally coupled to the gravity.
By Weyl rescaling the metric $g^{\rm E}_{\mu\nu}=\Phi g^{\rm J}_{\mu\nu}$ the Lagrangian in the Einstein frame reads
\begin{align}\label{eqn:LE}
{\SC L}=\sqrt{-g}\left[
\half{\SC R}-\half G_{ab}g^{\mu\nu}\partial_\mu\phi^{a}\partial_\nu\phi^{b}-V(\phi^I)\right],
\end{align}
where $\phi^{a}=(s,h)$ and $a=1,2$.
The scalar potential is
$V=\Phi^{-2}V_{\rm J}$.
In the Einstein frame the kinetic term for the scalar fields involves nontrivial field space metric
\begin{align}\label{eqn:FSmetric_cubic}
G_{ss}=\ &\frac{
(1+\xi h^2)\kappa +\frac{1}{24} (\omega^2+2\zeta) s^4}{\Phi^2},\nn\\
G_{sh}=\ &G_{hs}=-\frac{\xi hs(1-\frac{3}{2}\omega s-\zeta s^2)}{\Phi^2},\nn\\
G_{hh}=\ &\frac{6\xi^2h^2+\Phi}{\Phi^2}.
\end{align}
The Christoffel symbol for the field space is computed from the metric $G_{ab}$ as
\begin{align}
\Gamma^{s}_{ss}&=\frac{(\omega+2\zeta s) \big[s^2-12 (1+\xi h^2+6\xi^2 h^2)\big]
-(1-\omega s)\omega s^2}{12C}\crcr
&\quad+\frac{s(1-\frac 32 \omega s-\zeta s^{2})}{3\Phi}\,,\crcr
\Gamma^{s}_{sh}&=-\frac{\xi h}{\Phi}\,,\crcr
\Gamma^{s}_{hh}&=-\frac{(1+6\xi)(1-\frac 32 \omega s-\zeta s^2)s}{6C}\,,\crcr
\Gamma^{h}_{ss}&=-\frac{(\zeta+\half\omega^2)\xi h s^2}{C},\crcr
\Gamma^{h}_{sh}&=\frac{(1-\frac 32 \omega s-\zeta s^{2})s}{6\Phi}\,,\crcr
\Gamma^{h}_{hh}&=\frac{12(1-\xi h^{2})-2(1-\omega s)s^2+\zeta s^4}{12h\Phi}\nn\\
&\quad+\frac{12(2\omega s-1)+(24-s^2)\zeta s^2-\half\omega^2 s^4}{12hC}\,,
\label{eqn:CubicGamma}
\end{align}
where
\begin{align}
\xi\equiv\frac{\gamma}{4}-\sixth,
\end{align}
and
\begin{align}
C & \equiv  \Phi^{3} \mathrm{det} \, G_{ab}\,\crcr
&= (1+\xi h^2+6\xi^2 h^2)\kappa+\frac{(2\zeta+\omega^2)s^4}{24}\,.
\end{align}
The scalar curvature of the field space is
\begin{align}
R=-\frac{1}{3}-\frac{\Phi^{2}}{6C^{2}}(1+6\xi)(2\zeta+\omega^2)s^2\,.
\label{eqn:Ricciscalar_cubic}
\end{align}
In two dimensions the Riemann and the Ricci curvature tensors
are written using the scalar curvature as
$R^a{}_{bcd}=\half R(\delta^a_c G_{bd}-\delta^a_d G_{bc})$ and
$R_{ab}=\half R G_{ab}$.\\\\

\begin{figure*}[t]
\includegraphics[width=59mm]{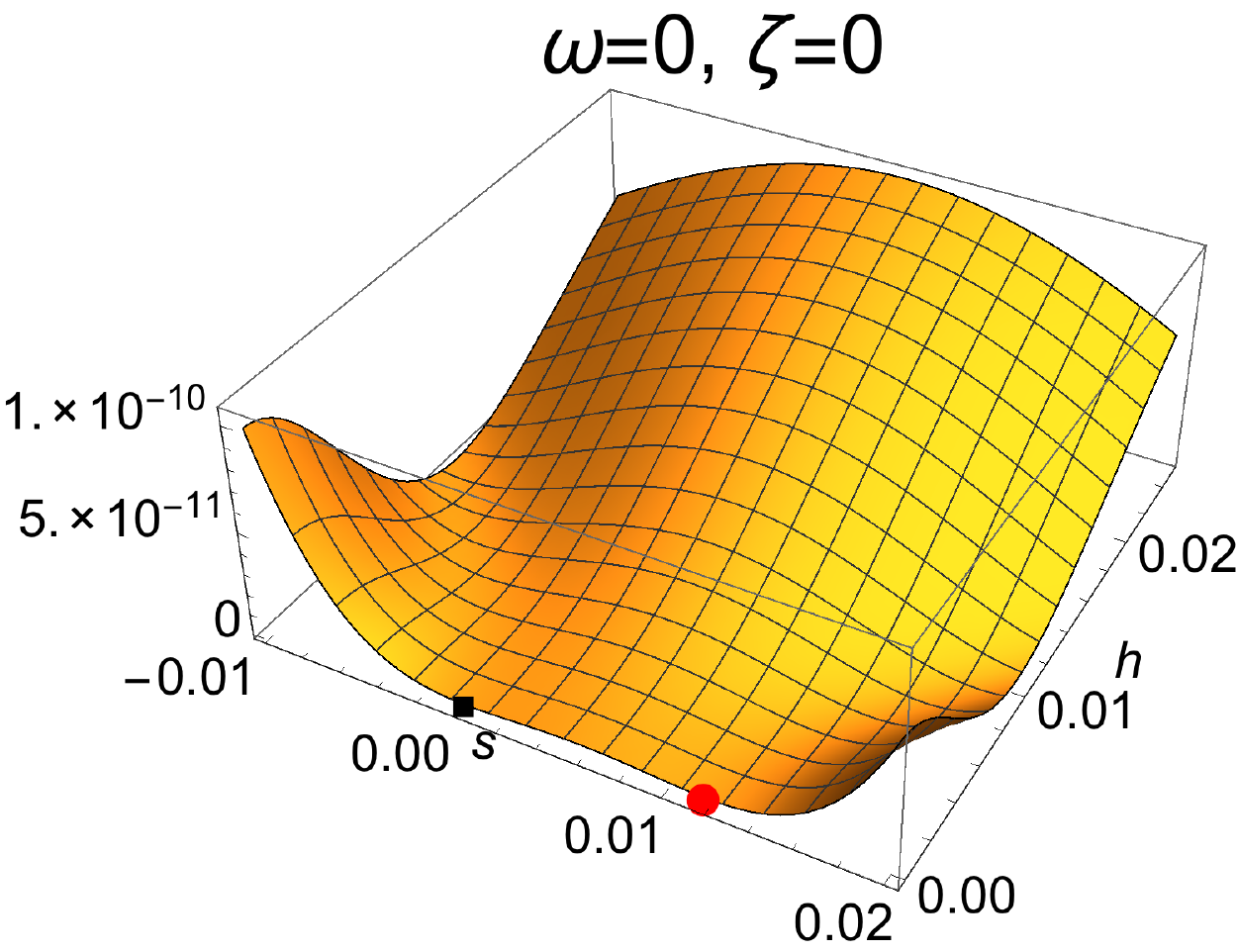}
\includegraphics[width=59mm]{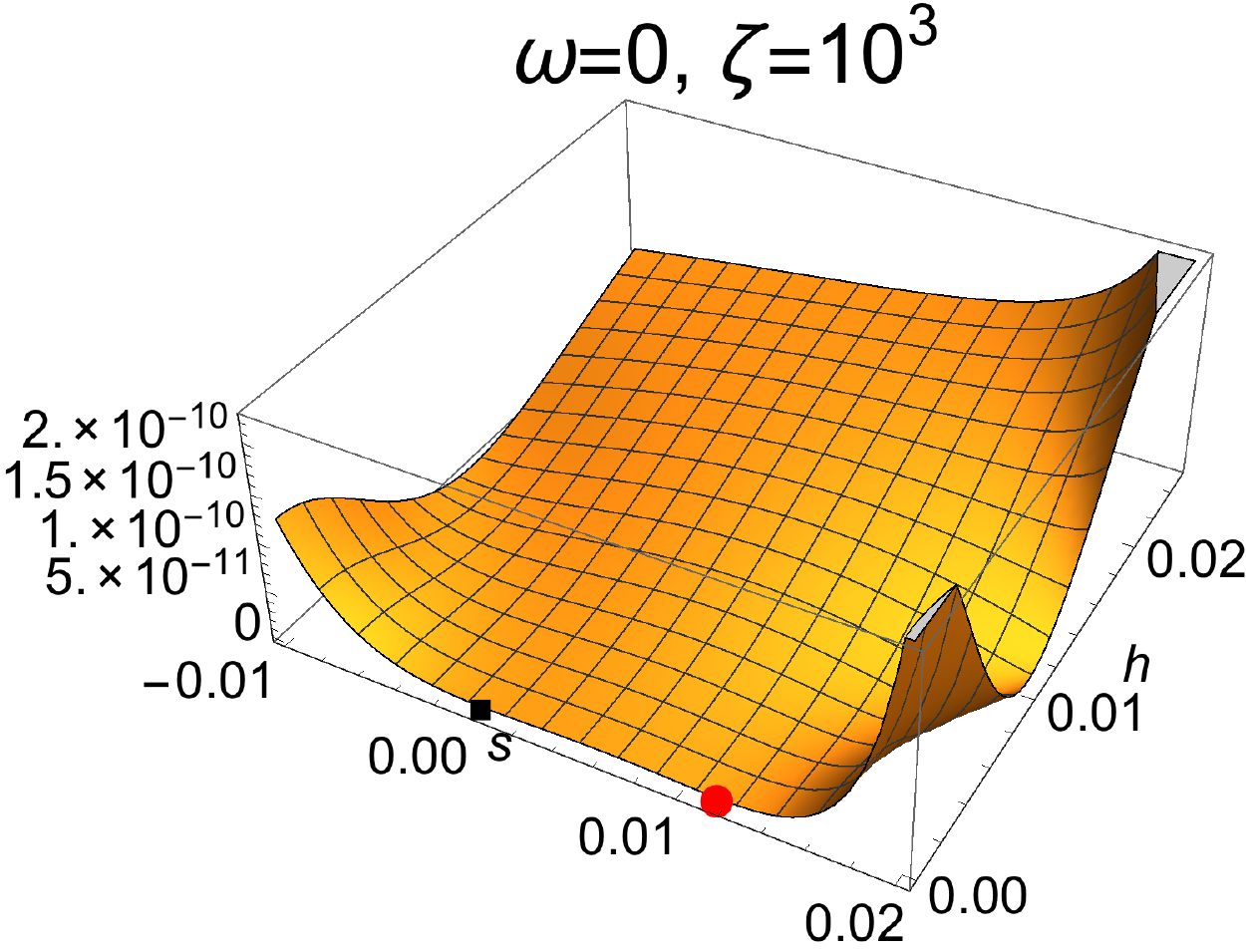}
\includegraphics[width=59mm]{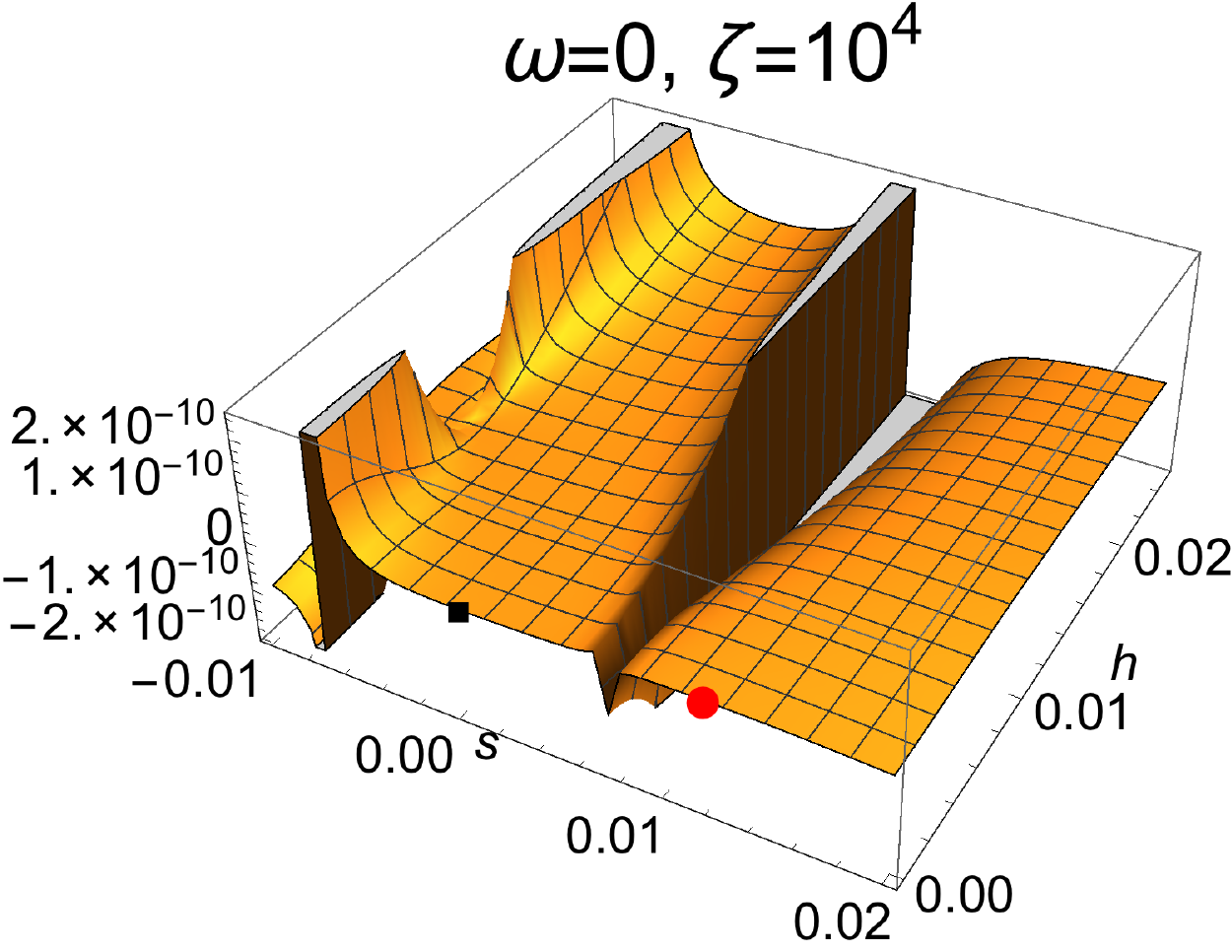}
\caption{\label{fig:stability}
The shape of the potential for $\zeta=0$ (left), $\zeta=10^3$ (center), and $\zeta=10^4$ (right) when the $\omega$ parameter is fixed to zero. The black square and the red circle are respectively the GUT vacuum and the SM vacuum. The $\xi$ parameter is chosen to be $\xi=5285$, which yields the Planck-normalized scalar power spectrum when $\omega=-116$ and $\zeta=10^4$.
}
\end{figure*}

\subsection{The sextic K\"{a}hler model}\label{sec:sextic}

In the above setup we included the noncanonical term proportional to $\widetilde\omega$ in the
K\"{a}hler potential \eqref{eqn:cubicPhi1}.
The effect of this term is to enlarge the parameter space so that the inflaton trajectory is allowed to terminate at the SM vacuum $(s, h)=(v,0)$ \cite{Arai:2011nq}.
For the same purpose we may alternatively consider the following K\"{a}hler potential:
\begin{align}\label{eqn:sexticPhi1}
\Phi &=1-\third\left(
\Tr\Sigma^\dag\Sigma + |H|^2+|\overline H|^2+\cdots\right)\crcr
&\quad
-\frac{\gamma}{2}\left(
\overline H H+H^\dag \overline H^\dag\right)\crcr
&\quad
+\frac{\zeta}{3}\left(
\Tr\Sigma^\dag\Sigma\right)^2
+\frac{\widetilde\omega}{3}\left(
\Tr\Sigma^\dag\Sigma\right)^3.
\end{align}
The term proportional to $\widetilde\omega$ gives a sextic term in $S$,
\begin{align}\label{eqn:sexticPhi2}
\Phi &=1-\third\left(
|S|^2+|H_u^0|^2+|H_d^0|^2\right)
+\frac{\gamma}{2}\left(H_u^0H_d^0+\text{c.c.}\right)\crcr
&\quad
+\frac{\zeta}{3}|S|^4
+\frac{\widetilde\omega}{3}\left|S\right|^6.
\end{align}
We take this as our second option of the K\"{a}hler potential that will be used in the supergravity embedding of the $SU(5)$ GUT model.

\subsubsection{Jordan frame}\label{sec:sexticJordan}
Using the same parametrization of the D-flat direction along $H_u$-$H_d$
and the same normalization of the $S$-field, we find that the Lagrangian
in the Jordan frame takes the same form \eqref{eqn:LJ}, but now with
%
\begin{align}
\Phi &=1-\sixth s^2+\frac{\zeta}{12} s^4+\frac{\omega}{6} s^6 +\xi h^2,\\
\kappa &\equiv 1-2\zeta s^2-9\omega s^4,\label{eqn:kappasextic}
\end{align}
where we have used
\begin{align}
\omega\equiv\frac{\widetilde\omega}{4},
\qquad
v\equiv\sqrt 2\ \widetilde v.
\end{align}
The scalar potential in the Jordan frame reads
\begin{widetext}
\begin{align}\label{eqn:VJsextic}
V_{\rm J}
=
\frac{3}{40}\left\{
\rho^2(s-v)^2h^2
+\frac{1}{\kappa}\left[\frac{\rho}{2}h^2-\frac{\lambda}{3} s(s-v)\right]^2\right\}
-\frac{\left\{
\frac{2(\zeta +6\omega s^2)}{\kappa}\!\left[\frac{\rho}{2}h^2-
\frac{\lambda}{3}s(s-v)\right]\! s^3\!
+\rho v h^2
-\frac{\lambda}{3}v s^2
-3\gamma\rho h^2 (s-v)\right\}^2}
{80\left[4+\gamma(\frac 32 \gamma-1) h^2+\frac{\zeta+8\omega s^2-\zeta\omega s^4}{3\kappa} s^4\right]}.
\end{align}
\end{widetext}

\subsubsection{Einstein frame}\label{sec:sexticEinstein}

After the Weyl transformation of the spacetime metric
$g^{\rm E}_{\mu\nu}=\Phi g^{\rm J}_{\mu\nu}$
the Lagrangian in the Einstein frame is written in the form of \eqref{eqn:LE}, with the scalar potential
\begin{align}
V=\Phi^{-2}V_{\rm J}
\end{align}
using \eqref{eqn:VJsextic} and the metric of the field space
\begin{align}\label{eqn:FSmetric_sextic}
G_{ss} &=\frac{
(1+\xi h^2)\kappa+\frac{1}{12}(\zeta+8\omega s^2-\zeta\omega s^4) s^4}{\Phi^2},\nn\\
G_{sh} &=\ G_{hs}=-\frac{\xi hs(1-\zeta s^2-3\omega s^4)}{\Phi^2},\nn\\
G_{hh} &=\ \frac{6\xi^2h^2+\Phi}{\Phi^2}.
\end{align}
The Christoffel symbol of the field space is
\begin{align}\label{eqn:SexticGamma}
\Gamma^{s}_{ss}&=
\frac{
(1-\zeta s^2-3\omega s^4)s}{3\Phi}\crcr
&\hspace{-10mm}+
\frac{
(\zeta +12\omega s^2-2\omega\zeta s^4) s^3
-12 s(1+\xi h^2+6\xi^2 h^2)(\zeta+9\omega s^2)}{6C}\,,\cr
%
\Gamma^{s}_{sh}&=-\frac{\xi h}{\Phi}\,,\cr
%
\Gamma^{s}_{hh}&=-\frac{(1+6\xi)(1-\zeta s^2-3\omega s^4)s}{6C}\,,\cr
%
\Gamma^{h}_{ss}&=-\frac{\xi h s^2 (\zeta+12\omega s^2-3\zeta\omega s^4)}{C}\,,\cr
%
\Gamma^{h}_{sh}&=\frac{(1-\zeta s^2-3\omega s^4)s}{6\Phi}\,,\cr
%
\Gamma^{h}_{hh}&=\frac{12(1-\xi h^{2})-2s^2+\zeta s^4+2\omega s^6}{12h\Phi}\crcr
&+\frac{-12+(24-s^2)\zeta s^2+\omega s^4(108-8s^2+\zeta s^4)}{12hC}\,,
\end{align}
where
\begin{align}
C &\equiv \Phi^{3} \mathrm{det} G_{ab}\,\nonumber\\
&=(1+\xi h^2+6\xi^2 h^2)\kappa
+\frac{(\zeta+8\omega s^2-\omega\zeta s^4)s^4}{12}\,.
\end{align}
The scalar curvature of the field space is
\begin{align}
R=-\frac{1}{3}
-\frac{\Phi^{2}}{3C^{2}}
(1+6\xi)(\zeta+12\omega s^2-3\omega\zeta s^4)s^2
\,.
\label{eqn:Ricciscalar_sextic}
\end{align}

\subsection{Higgs inflation in $SU(5)$ GUT}\label{sec:HIinGUT}

We have seen above that assuming supergravity embedding with the K\"{a}hler potential either in the form of \eqref{eqn:cubicPhi1} or \eqref{eqn:sexticPhi1},
the $SU(5)$ GUT model with the superpotential \eqref{eqn:W}
leads to a system of two scalar fields described by the Lagrangian \eqref{eqn:LE}.
Note that in the limit of trivial $s$-field dynamics, that is, if we set $s=v=0$,
the Jordan frame Lagrangian \eqref{eqn:LJ} becomes
\begin{align}
{\SC L}_{\rm J}=&\sqrt{-g_{\rm J}}\left[
\half\Phi{\SC R}_{\rm J}
-\half g_{\rm J}^{\mu\nu}\partial_\mu h\partial_\nu h
-\frac{3}{160}\rho^2 h^4\right],
\end{align}
with $\Phi=1+\xi h^2$.
This is the Lagrangian of the nonminimally coupled single field inflation model with a quartic self coupling, which has attracted much attention recently.
This model predicts a small tensor-to-scalar ratio compatible with the Planck and the WMAP observations; see e.g. \cite{Okada:2010jf}.
Since the field $h$ above is identified as (the D-flat component of) the SM Higgs field, this model is considered as a realization of the Bezrukov-Shaposhnikov scenario of SM Higgs inflation \cite{CervantesCota:1995tz,Bezrukov:2007ep} within supersymmetric $SU(5)$ grand unification.
The $s$ field is a component of the GUT Higgs field and, since the GUT symmetry is broken in our world, phenomenological consistency does not allow the single-field limit $s=v=0$.
Hence an honest multifield analysis is mandatory if we are to make prediction based on this model.
In the next section we present the results of numerical study of the multifield inflationary dynamics.
The technicalities of the formalism we use are summarized in Appendix.~\ref{sec:dynamics}.

\begin{figure*}
\includegraphics[width=44mm]{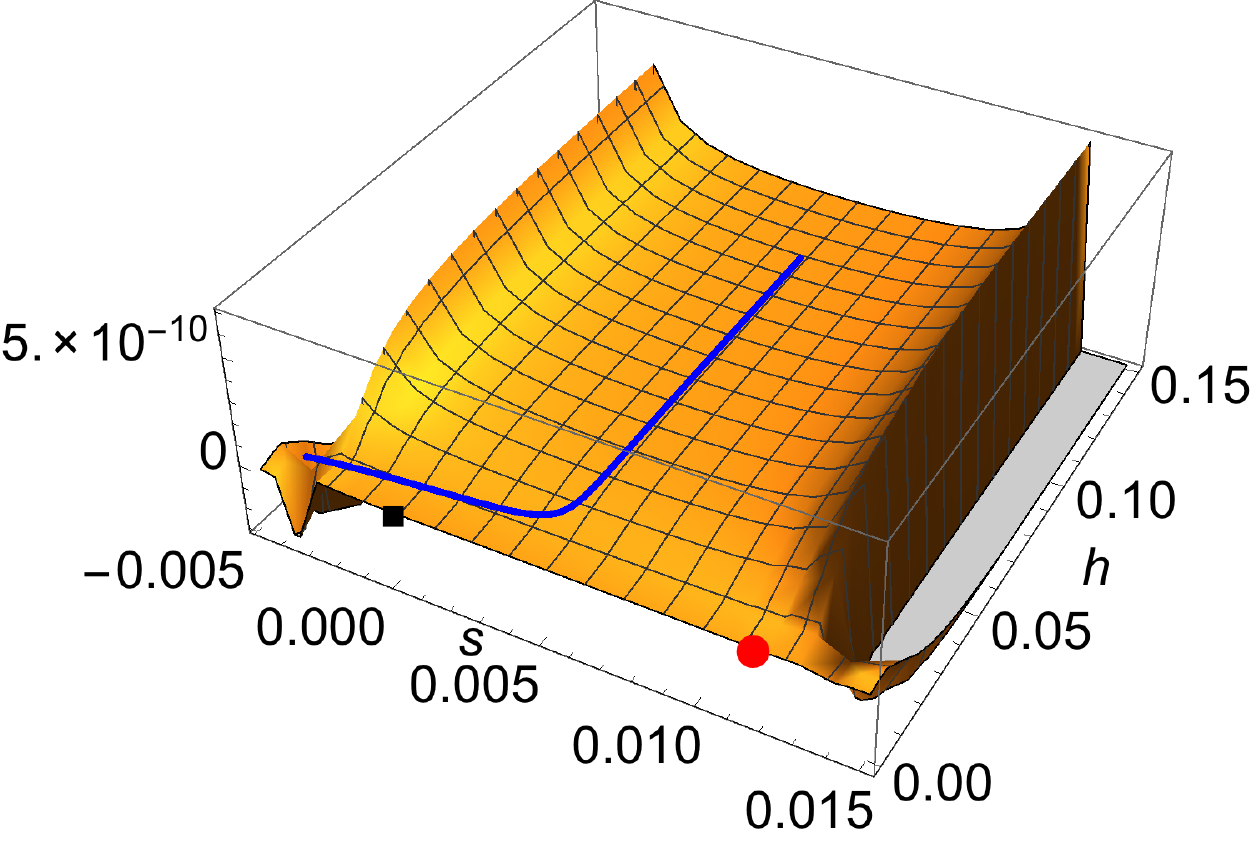}
\includegraphics[width=44mm]{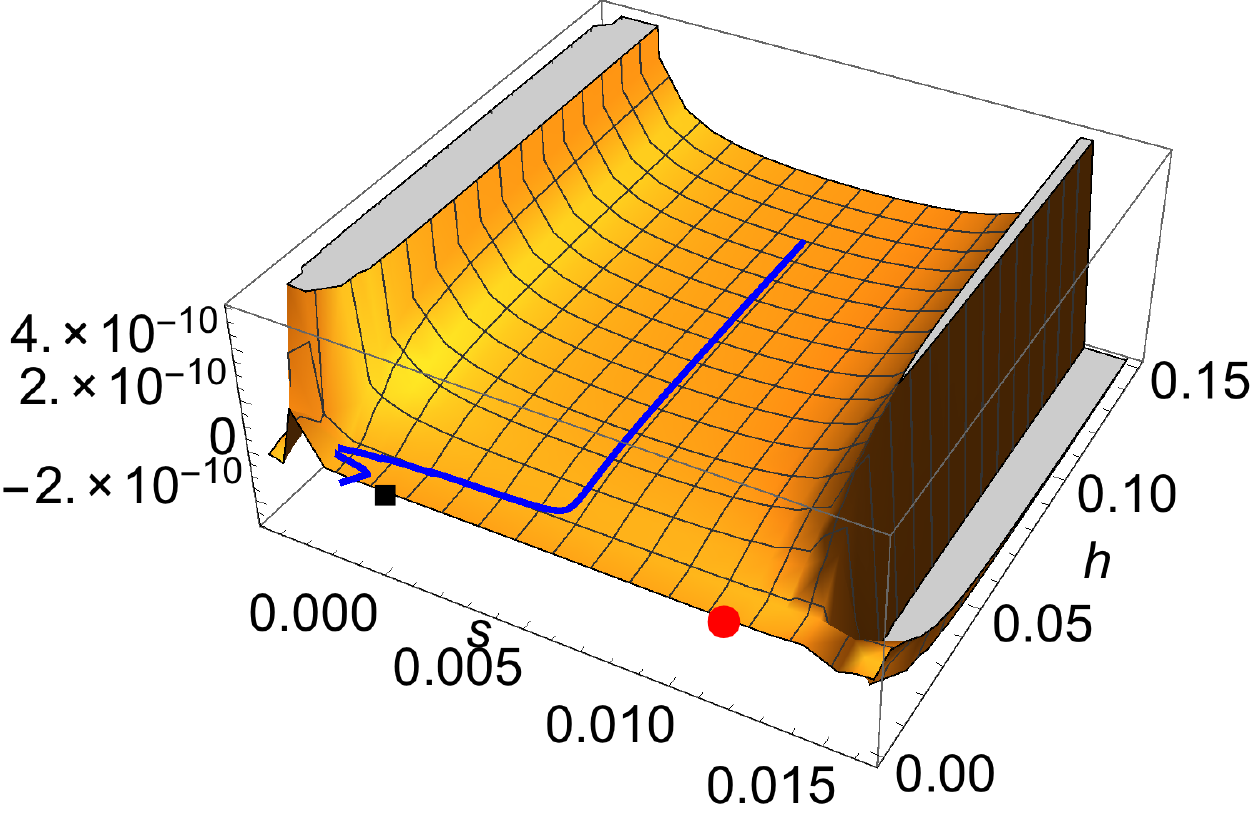}
\includegraphics[width=44mm]{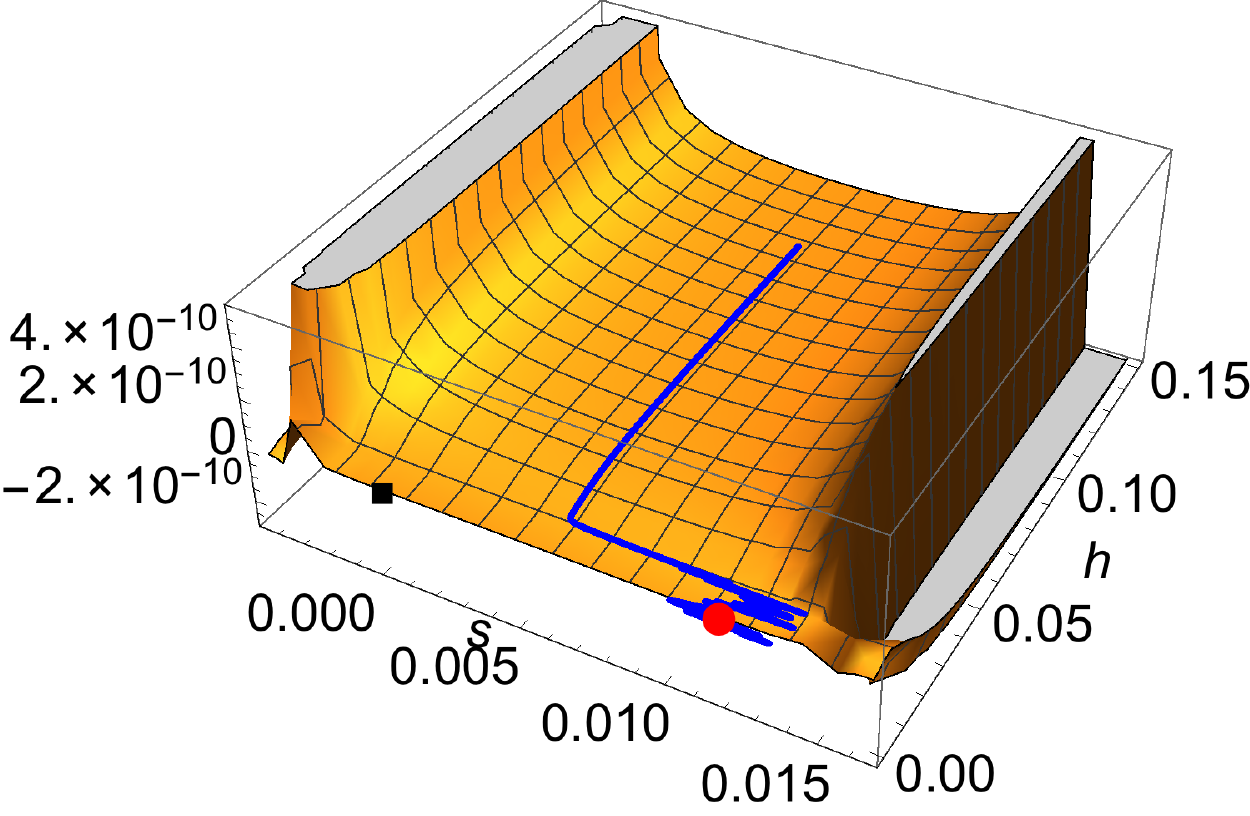}
\includegraphics[width=44mm]{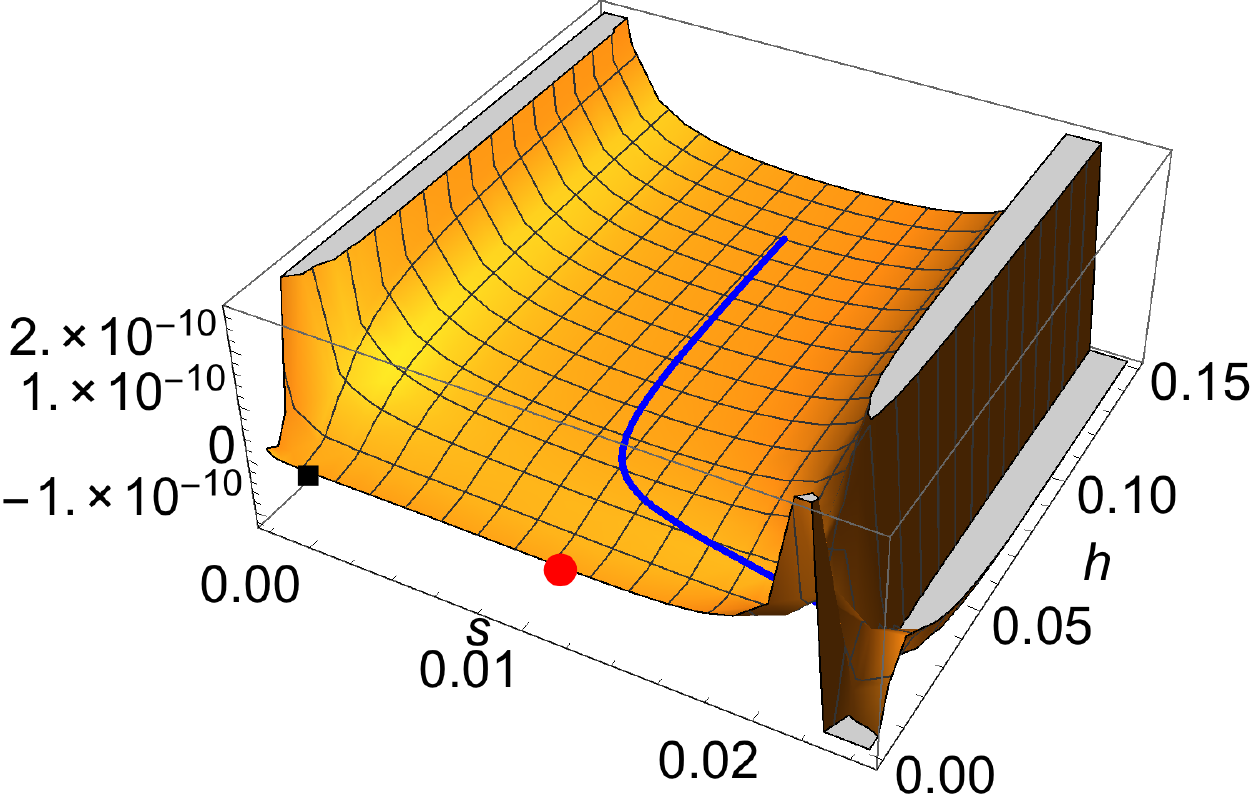}
\caption{\label{fig:cubic_traj}
Four types of inflaton trajectories in the cubic K\"{a}hler model found numerically:
(a) escape through the left wall, (b) settle in the GUT vacuum, (c) settle in the SM vacuum, (d) escape through the right wall from the left.
The $\omega$ parameter for (a), (b), (c) and (d) are respectively
$\omega=-100$, $-114$, $-116$ and $-200$.
The parameter $\zeta$ is chosen to be $\zeta=10^4$ for all cases.
}
\end{figure*}

\begin{figure*}
\includegraphics[width=80mm]{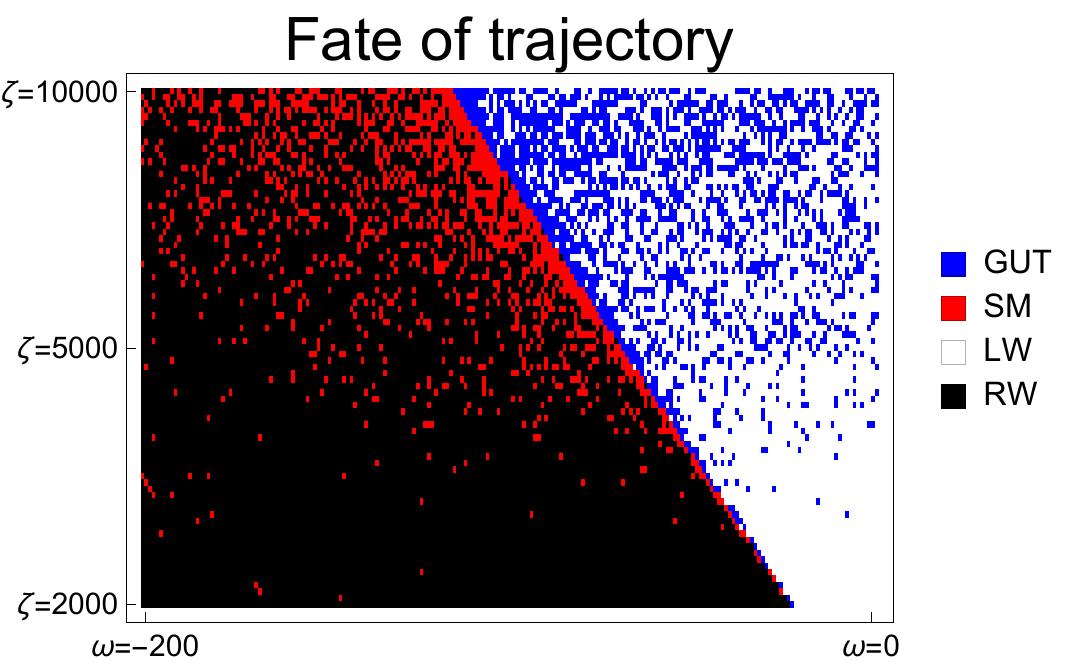}
\includegraphics[width=80mm]{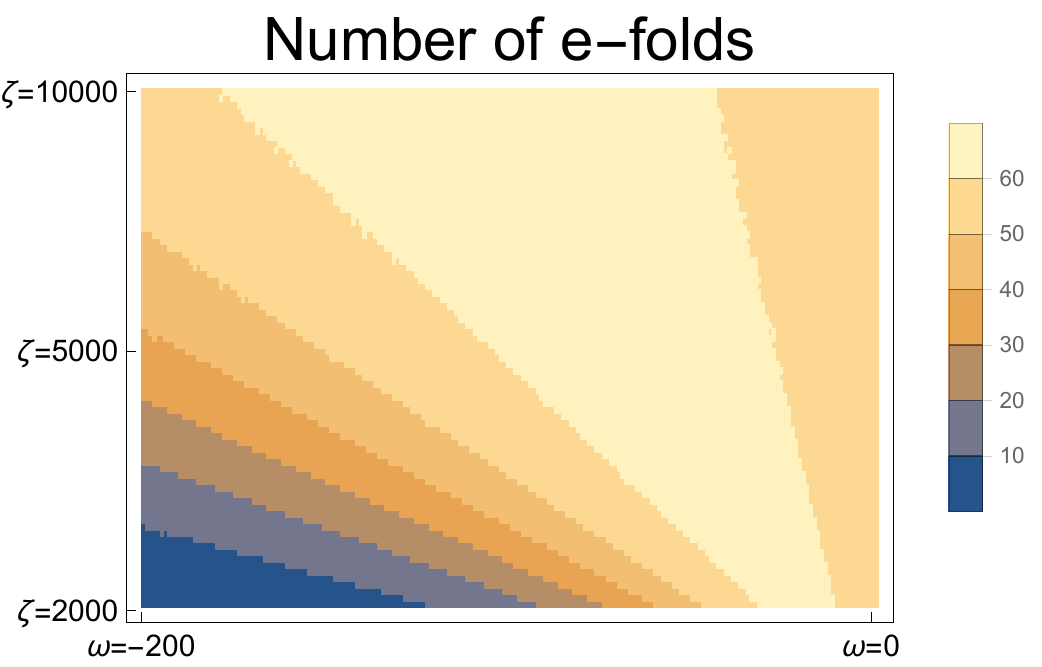}
\caption{\label{fig:cubic_solutions}
The four types of numerical solutions with different fates (left) and the
e-folding number (right) for numerical solutions in the parameter range
$2000\leq\zeta\leq 10000$ and $-200\leq\omega\leq 0$.
These parameters are changed with the stepsize $\Delta\zeta=100$, $\Delta\omega=1$.
}
\end{figure*}

\begin{table*}[t]
\begin{center}\begin{tabular}{cc||cccccccc}
$\zeta$ & $\omega$ & $h_*$ & $h_{{\rm end}}$ & $N_e$ & ${\C P}_{\C R}$ & ${\C P}_T$ & $n_s$ & $r$ & $f_{\rm NL}^{{\rm local}}$\\ \hline
10000 & -117 & 0.1174 & 0.0093 & 63.06 & 2.134$\times 10^{-9}$ & 6.570$\times 10^{-12}$ & 0.9670 & 0.0031 & -1.105 \\
8000 & -94 & 0.1174 & 0.0096 & 63.04 & 2.318$\times 10^{-9}$ & 7.130$\times 10^{-12}$ & 0.9670 & 0.0031 & -1.271 \\
6000 & -71 & 0.1175 & 0.0102 & 63.00 & 2.537$\times 10^{-9}$ & 7.793$\times 10^{-12}$ & 0.9671 & 0.0031 & -3.527 \\
4000 & -48 & 0.1176 & 0.0117 & 62.87 & 2.808$\times 10^{-9}$ & 8.593$\times 10^{-12}$ & 0.9671 & 0.0031 & -5.341 \\
2000 & -25 & 0.1181 & 0.0155 & 62.35 & 3.180$\times 10^{-9}$ & 9.574$\times 10^{-12}$ & 0.9674 & 0.0030 & -6.894
\end{tabular}
\caption{
The values of the field $h$ at the horizon crossing $h_{*}$ and at the end of slow roll $h_{\rm end}$, the e-folding number $N_e$, the scalar and tensor power spectra ${\C P}_{\C R}$ and ${\C P}_T$, the scalar spectral index $n_s$, the tensor-to-scalar ratio $r$ and the local-type nonlinearity parameter $f_{\rm NL}^{{\rm local}}$ in the cubic K\"{a}hler model as the parameters $\zeta$ and $\omega$ are varied.
The initial value of the $h$ field is chosen to be $h_{\rm init}=0.12$ and the parameter $\xi$ is fixed to 5285 using the Planck normalization of the scalar power spectrum when $(\zeta,\omega)=(10000,-116)$ and e-foldings 60.
The $N_e$ in the table is the e-folding number between $h_{\rm init}$ and $h_{{\rm end}}$.
\label{tab:cubic}}
\end{center}
\end{table*}

\begin{figure*}[t]
\includegraphics[width=59mm]{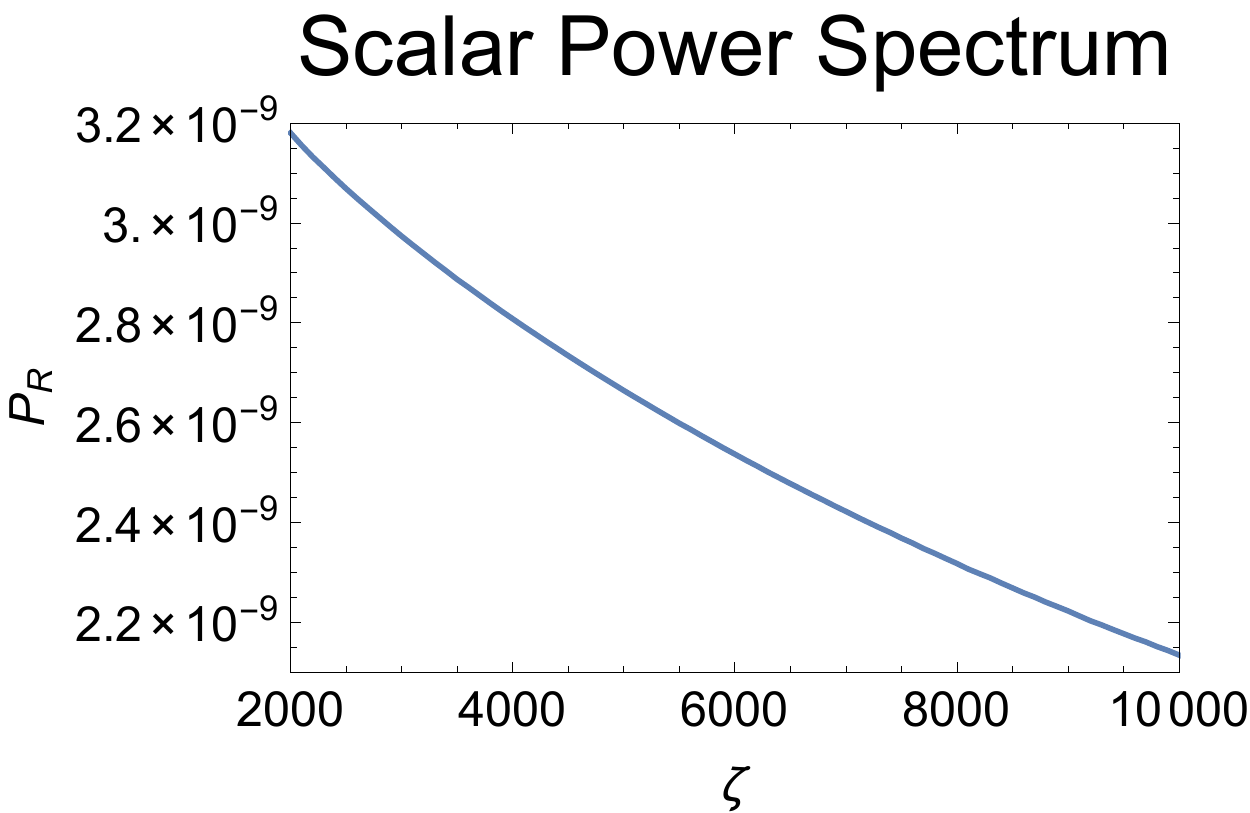}
\includegraphics[width=59mm]{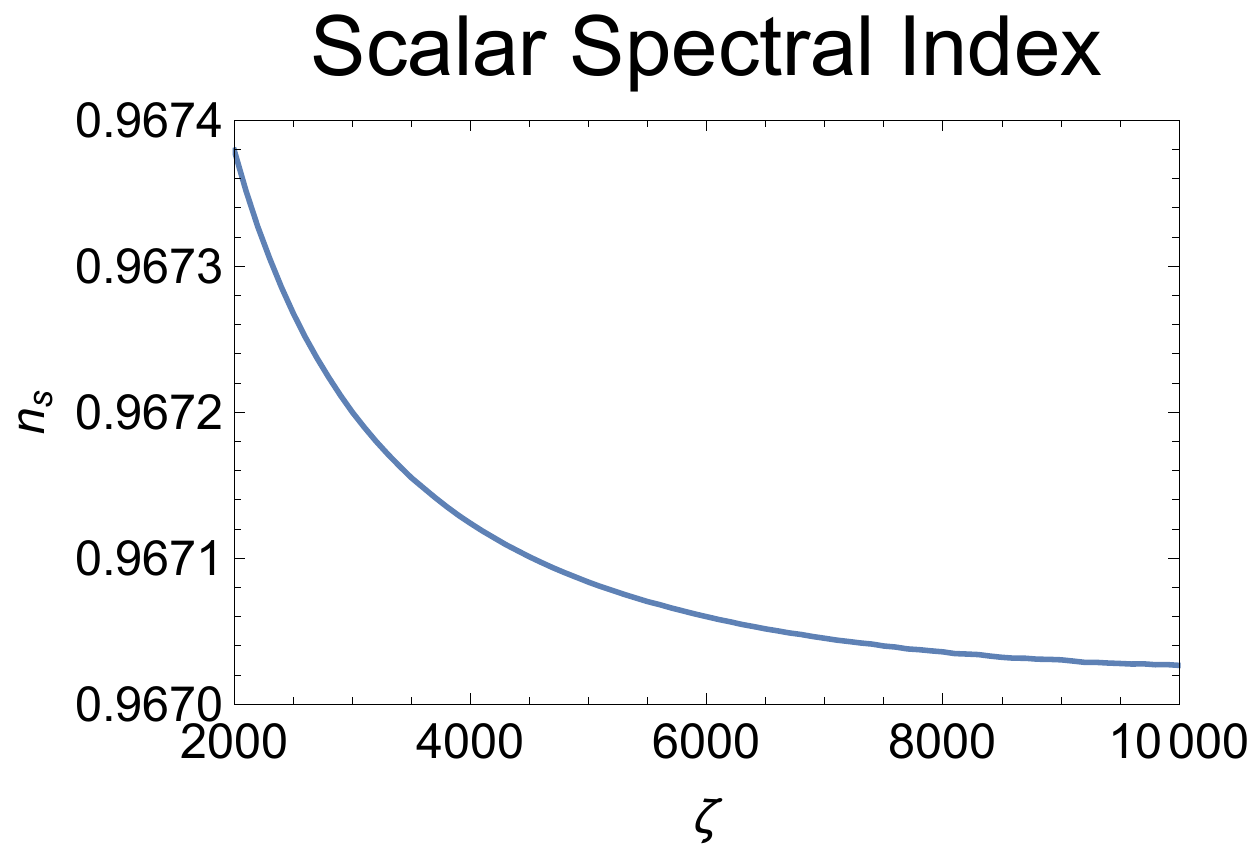}
\includegraphics[width=59mm]{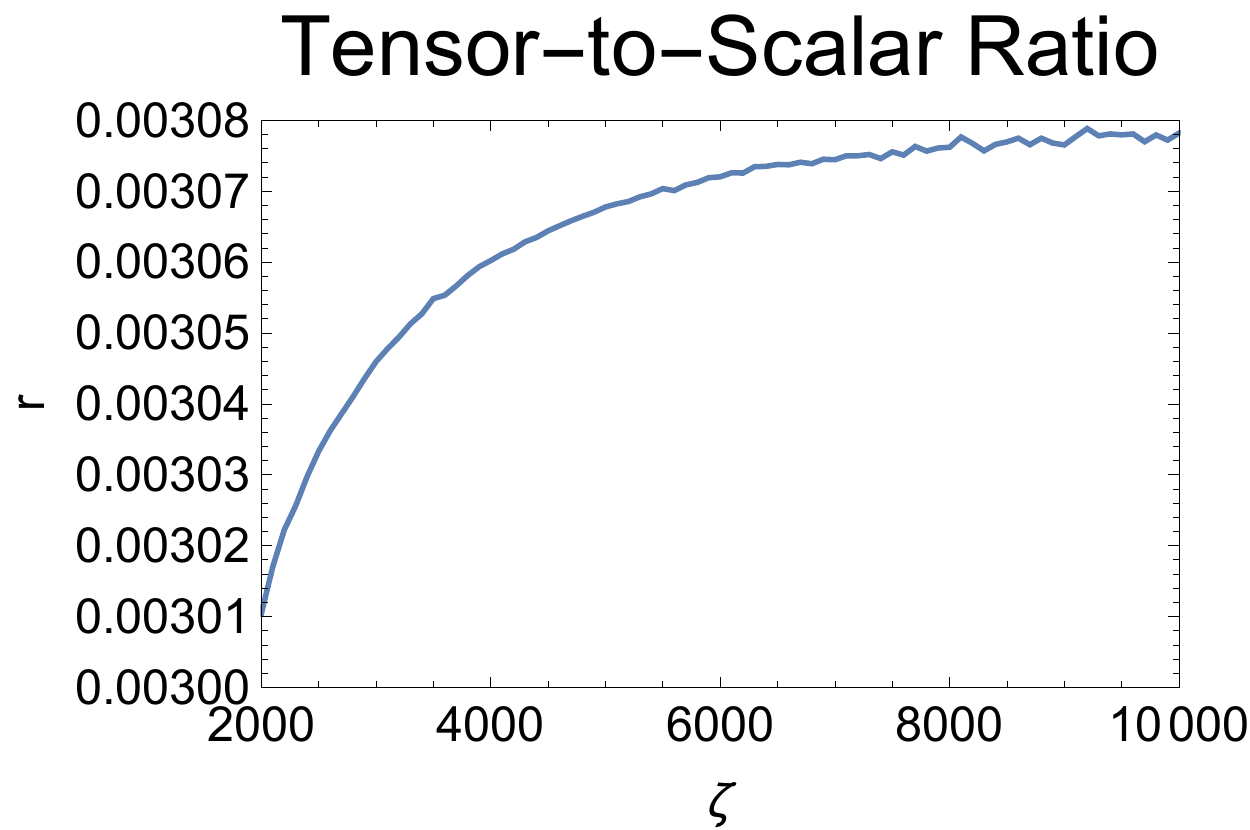}
\caption{\label{fig:cubic_CO}
The scalar power spectrum, the scalar spectral index and the tensor-to-scalar ratio for the cubic K\"{a}hler model.
}
\end{figure*}

\section{Numerical results}\label{sec:numerics}

In this section we discuss the multifield dynamics of the inflationary model introduced in the previous section.
We first comment on the shape of the inflaton potential when
$\omega=\widetilde\omega=0$ in \eqref{eqn:cubicPhi1} or \eqref{eqn:sexticPhi1} (in the $\omega=0$ limit these two models are identical).
The potential in this case is not phenomenologically viable as the SM vacuum cannot be reached after inflation.
We then investigate the cases for nonzero $\omega$, first in the presence of the cubic term \eqref{eqn:cubicPhi1} and then the sextic term \eqref{eqn:sexticPhi1} of $S$ in the K\"{a}hler potential;
we will see that phenomenologically viable inflaton trajectories are allowed in both cases.
Next, cosmological parameters, including the scalar spectral index, the tensor-to-scalar ratio, the isocurvature fraction and non-Gaussianity are computed.
The formalism we use in our numerical code\footnote{
To crosscheck and debug our numerical code we used TransportMethod \cite{Dias:2015rca} and 
MultiModeCode \cite{Price:2014xpa}.
}
is summarized in Appendix. \ref{sec:dynamics}.
For computation of the non-Gaussianities we use another set of numerical code developed in \cite{Kim:2014vsa, Kawai:2014gqa}.

\subsection{Inflaton potential of the $SU(5)$ GUT model}

The inflationary model we study here includes 5 tunable parameters:
$\rho$, $\lambda$, $\xi$, $\omega$, and $\zeta$.
The first two of them concern the physics of grand unification and are ${\C O}(1)$.
For the sake of concreteness we shall set them to $\rho = \lambda = 0.5$ in the following analysis.
The inflationary dynamics is not very sensitive to the values of $\rho$ and $\lambda$ \cite{Arai:2011nq}.
The parameter $\xi$ corresponds to the nonminimal coupling in the case of the SM
Higgs inflation model and is crucial for the slow-roll dynamics.
We fix this parameter by the Planck normalization of the fluctuation amplitude \cite{Ade:2015xua,Ade:2015lrj}
\begin{align}\label{eqn:PlanckNorm}
A_s=2.207 \times 10^{-9}
\quad
\text{(TT, TE, EE $+$ low P)}.
\end{align}
The number of e-folds between the horizon crossing of the cosmic microwave background (CMB) scale and the end of inflation is chosen to be $N_e=60$.
For a given inflaton trajectory, the end of inflation is characterized by the condition $\epsilon=1$, with the slow roll paremeter $\epsilon$ defined in \eqref{eqn:epsilon}.
Integrating the Hubble parameter backward in time from there along the inflaton trajectory for $N_e=60$ we locate the horizon crossing of the CMB scale.
Solving the evolution of the fluctuations forward in time from there we fix the $\xi$ parameter by the condition that the adiabatic mode at the end of inflation is normalized by the Planck observation \eqref{eqn:PlanckNorm}.
Note that in multifield inflation the amplitude of the adiabatic mode at the end of inflation may differ from the value at the horizon crossing, due to the isocurvature effects.

To see the effects of the remaining two parameters $\zeta$ and $\omega$, let us look at the shape of the potential (Fig.~\ref{fig:stability}, the left panel) when $\zeta=\omega=0$.
As our focus is on Higgs inflation realized in supersymmetric GUT, we are interested in the inflaton trajectory that lies along the direction of $h$.
However, the potential is seen to exhibit tachyonic instability in the direction of $s$ field and hence slow roll in the $h$ direction will not take place.
The instability is removed by including a quartic term $\zeta |S|^4$ in the K\"{a}hler potential (Fig.~\ref{fig:stability}, the center and the right panel).
In general, such higher terms can exist in supergravity.
While a larger value of $\zeta$ renders the inflaton potential more stable, there exists an upper bound of $\zeta$ in our context, as the zeros of the K\"{a}hler metric $\kappa=1-2\zeta s^2$ introduce singularities of the potential at $s=\pm 1/\sqrt{2\zeta}$, beyond which the supergravity Lagrangian is unreliable.
As the SM vacuum lies at $s=v\sim M_{\rm GUT}$ in our model, a scenario of inflation that ends up in the SM vacuum requires $\zeta<1/2v^2$ ($\simeq 7\times 10^3$, assuming $v\simeq 2\times 10^{16}$ GeV).
Within this range of $\zeta$, no inflaton trajectories terminating in the SM vacuum can be found.
This problem may be solved by modifying the K\"{a}hler potential further, with the terms parametrized by
$\omega$ \cite{Arai:2011nq}.
In the following we study the two cases explained in Sec.~\ref{sec:model}, namely the model
\eqref{eqn:cubicPhi1} with an additional cubic $S$ term, and the one
\eqref{eqn:sexticPhi1} with an additional sextic $S$ term.

\subsection{The cubic K\"{a}hler model}

In this case the nontrivial component of the K\"{a}hler metric $\kappa$ is modified as \eqref{eqn:kappacubic}.
Its zeros are shifted to
$s=s_\pm\equiv(-\omega\pm\sqrt{2\zeta+\omega^2})/2\zeta$.
These are the location of the singularity walls that are used to tame the tachyonic instabilities.
It is easy to see that $s_\pm$ are real when $\zeta>-\omega^2/2$.
As the GUT vacuum and the SM vacuum must be both between the walls, 
we impose $s_-<0<v<s_+$, leading to $\zeta>0$ and $\omega<\frac{1}{2v}-\zeta v$.
The purpose of introducing the $\omega$ parameter has been to shift the walls so that the SM vacuum is favored over the GUT vacuum; this implies $\omega<0$.
Finally, $s_+-s_-\ll 1$ to cure the tachyonic instabilities, which gives to
$\zeta\gg 1$.

\begin{figure*}[t]
\includegraphics[width=59mm]{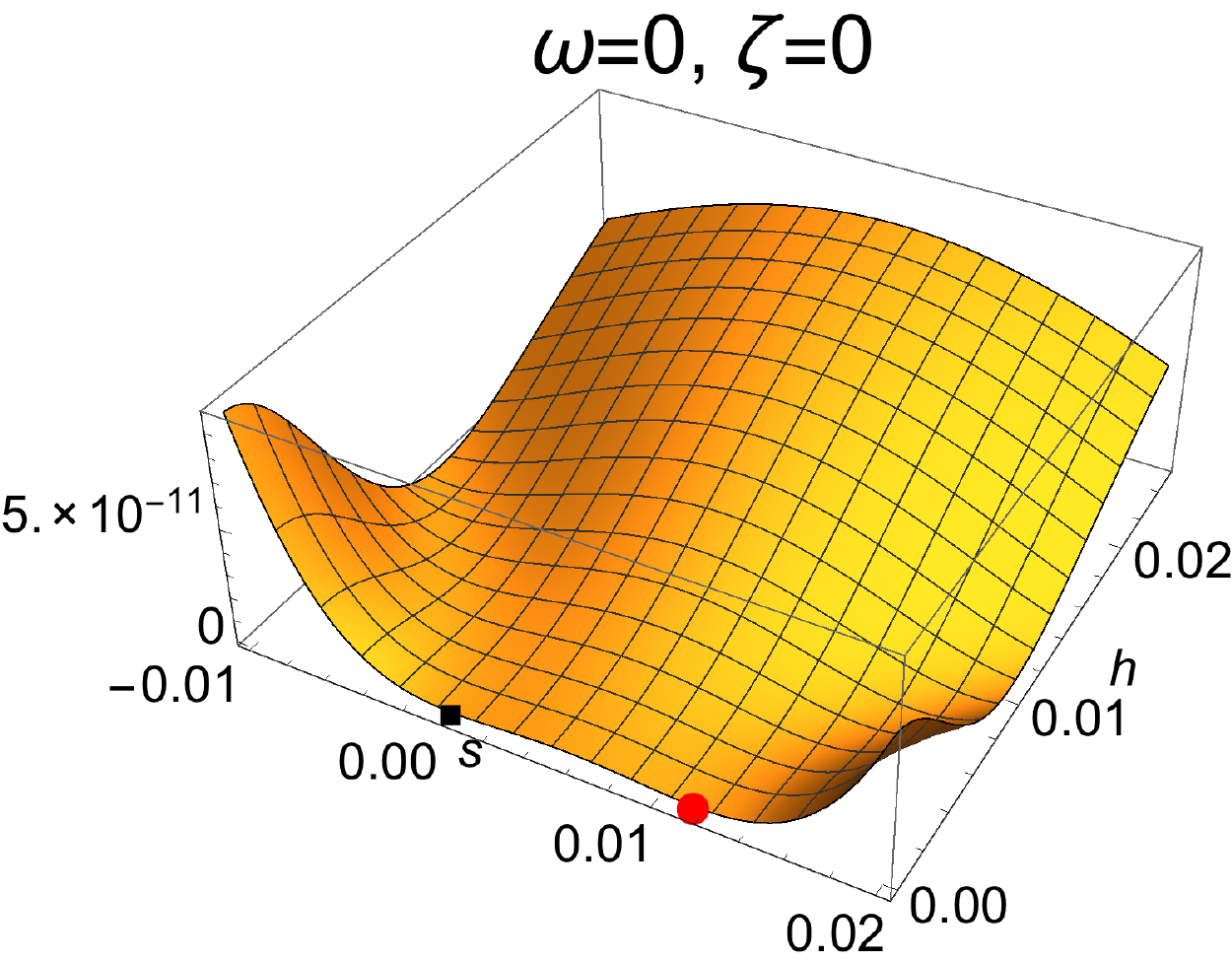}
\includegraphics[width=59mm]{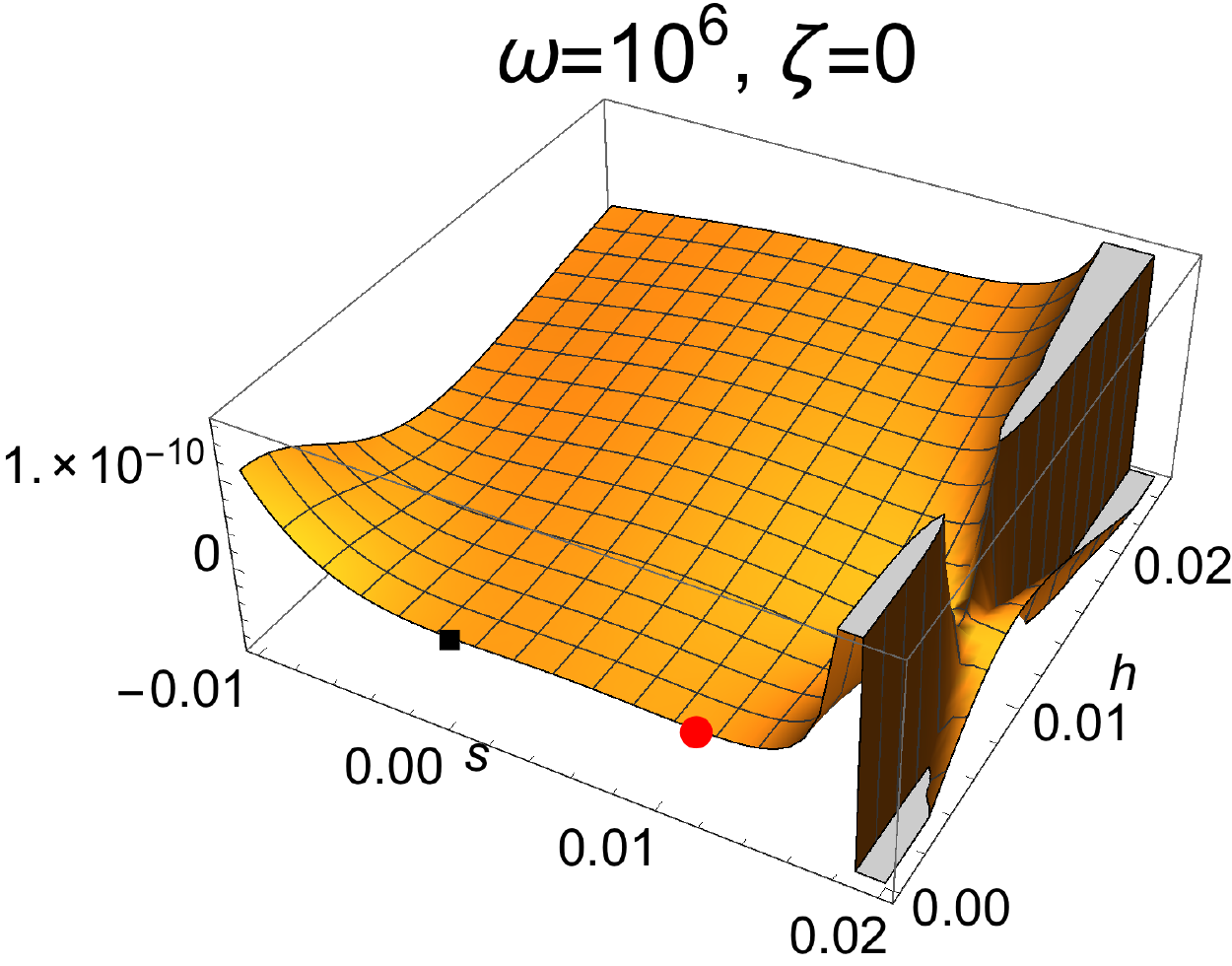}
\includegraphics[width=59mm]{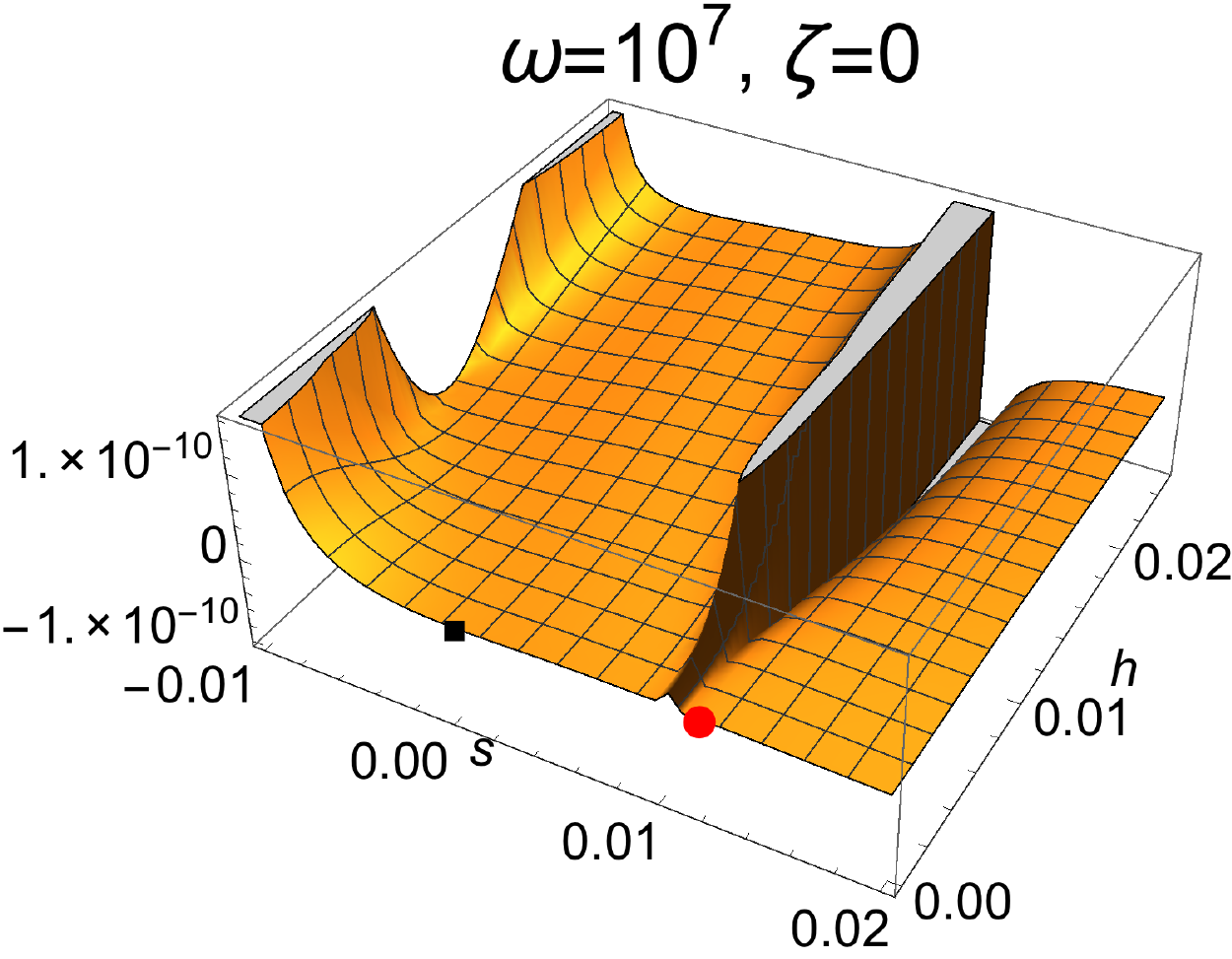}
\caption{\label{fig:sextic_stability}
The shape of the sextic K\"{a}hler potential for $\omega=0$ (left), $\omega=10^6$ (center), and $\omega=10^7$ (right) when the $\zeta$ parameter is fixed to zero. The black square and the red circle are respectively the GUT vacuum and the SM vacuum. The $\xi$ parameter is chosen to be $\xi=6450$, which yields the Planck-normalized scalar power spectrum when $\omega=10^{7}$ and $\zeta=-3000$.
}
\end{figure*}

\subsubsection{Background solutions}
\label{subsubsec:cubic_BG}

As our interest is in the Higgs inflation model realized in GUT, we focus on inflaton trajectories which are sufficiently straight along the SM Higgs $h$ direction in the large field region.
For this a large value of $\zeta$ is needed, and as an example of large enough $\zeta$ we
choose $\zeta=10000$.
As mentioned earlier, we fix the value of $\xi$ using the Planck normalization of the density fluctuations.
In doing so we solve the background equation fully numerically without using slow-roll approximation.
See Appendix \ref{subsec:BkgSol} and \ref{subsec:CosmoObs} for the details of the procedure.
For $\zeta=10000$ and $\omega=-116$ (the reason for choosing this value of $\omega$ is explained below), we find that the Planck normalization \eqref{eqn:PlanckNorm} gives
$\xi=5285$.
We use this trajectory as a reference case for the cubic K\"{a}hler model.

To study the behavior of background solutions in the vicinity of this reference trajectory, we first
vary the value of $\omega$, keeping the values of $\zeta$ and $\xi$ fixed.
We find four types of numerical solutions for the inflaton trajectories, as shown in Fig.~\ref{fig:cubic_traj}.
In the figure, the initial value of $h$ is chosen to be $h=h_{\rm init}=0.12$ (which yields more than 60 e-folds in the reference case).
The first type of trajectory, which we call the LW-type solutions, makes a turn after the slow roll and escapes through a hole in the $s=s_-$ wall, as depicted in Fig.~\ref{fig:cubic_traj}a.
The second type (the GUT-type) reaches the GUT vacuum after the slow roll, as shown in Fig.~\ref{fig:cubic_traj}b.
The third type, which we call the SM-type, is the phenomenologically viable one that reaches the SM vacuum $s=v$ after the slow roll, as shown in Fig.~\ref{fig:cubic_traj}c.
The last one, the RW-type, is similar to the first but escapes through the wall at $s=s_+$, as shown in Fig.~\ref{fig:cubic_traj}d.

We next change the value of the $\zeta$ parameter.
The left panel of Fig.~\ref{fig:cubic_solutions} shows how the four types of numerical solutions above are distributed when both $\zeta $ and $\omega$ are varied.
For the $\xi$ parameter we use the reference value $\xi=5285$ throughout.
The initial value for $h$ is the same as above, $h_{\rm init}=0.12$, and the initial value of $s$ is chosen at a local minimum of the potential along the $h=h_{\rm init}$ line.
The solutions of the LW-type and the GUT-type, as well as the solutions of the SM-type and the RW-type, are seen to be mixed.
In contrast, there is a clear line separating the solutions LW+GUT and the solutions SM+RW, which is found to be
\begin{align}\label{eqn:CubicSeparatrix}
\omega\approx -0.0115\,\zeta-1.00.
\end{align}
The right panel of Fig.~\ref{fig:cubic_solutions} shows the number of e-folds between $h=h_{\rm init}=0.12$ and $h=h_{\rm end}$ at the end of the slow roll, characterized by $\epsilon=1$.
Away from the separatrix \eqref{eqn:CubicSeparatrix}, the number of e-folds is seen to decrease; for such a trajectory, a larger value of $h_{\rm init}$ is required to solve the horizon problem.
The escape solutions LW and RW are numerical artifacts and should not be considered as (classical) physical solutions.
At the holes in the walls,
\begin{align}
\frac{\left[\half\rho h^2-\third\lambda s(s-v)\right]^2}{\kappa}
\end{align}
in the (Jordan frame) potential $V_{\rm J}$ \eqref{eqn:VJcubic} becomes indefinite.
These holes are thus located at the points
$(s,h)=(s_\pm, \sqrt{2\lambda s_\pm (s_\pm -v)/3\rho})$.
The walls are infinitely high except exactly at these points, and as these points are measure zero,
the LW and RW solutions should respectively become the GUT and the SM solutions if the step size is sent to infinitesimal.
Besides, the supergravity effective Lagrangian is unreliable near the singularity walls.

As emphasized, phenomenologically viable inflaton trajectories are those reaching the SM vacuum after the slow roll.
Since the trajectories that approach very close to the singularity walls are unreliable, we take
solutions in the vicinity of the line \eqref{eqn:CubicSeparatrix} as benchmark cases of the cubic K\"{a}hler model.
The left panel of Fig.~\ref{fig:cubic_solutions} shows that the GUT and SM solutions become sparse as $\zeta$ becomes small, indicating that obtaining a reliable numerical solution becomes increasingly difficult in this parameter region.
The parameter values for the reference trajectory $\xi=5285$ and $(\zeta,\omega)=(10000,-116)$ are chosen so that it is a SM solution near the separatrix \eqref{eqn:CubicSeparatrix} that does not approach too close to the singularity walls.
Note that away from this reference point, the $\xi$ parameter for the solutions in Fig.~\ref{fig:cubic_solutions} is not strictly Planck-normalized.

\begin{figure*}[t]
\includegraphics[width=44mm]{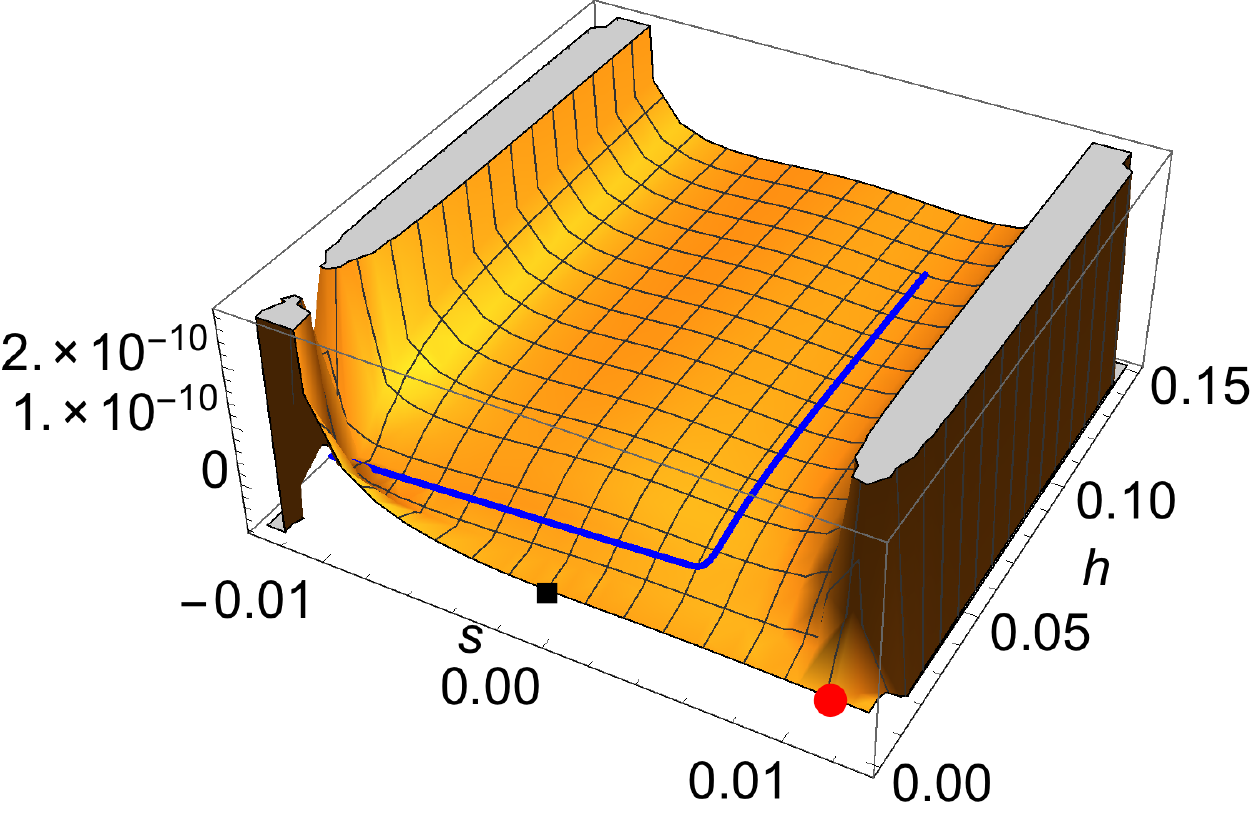}
\includegraphics[width=44mm]{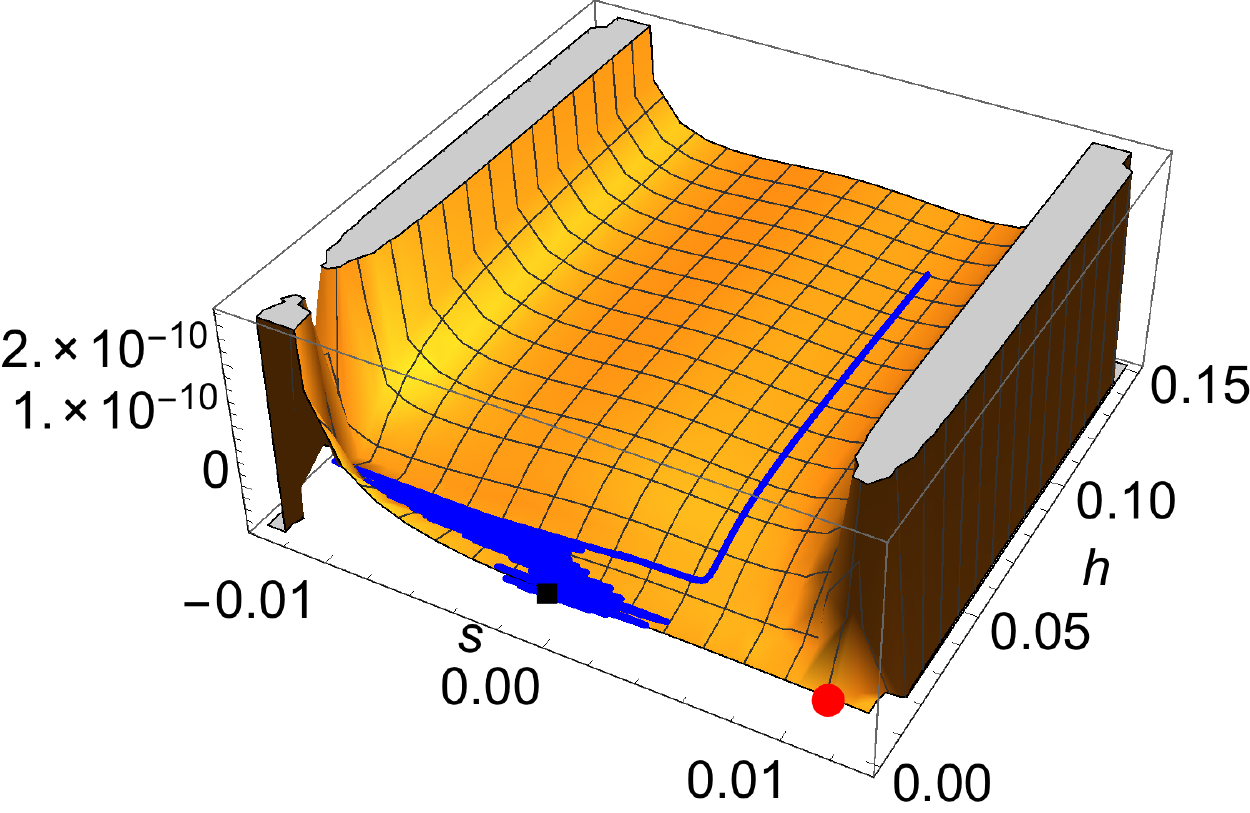}
\includegraphics[width=44mm]{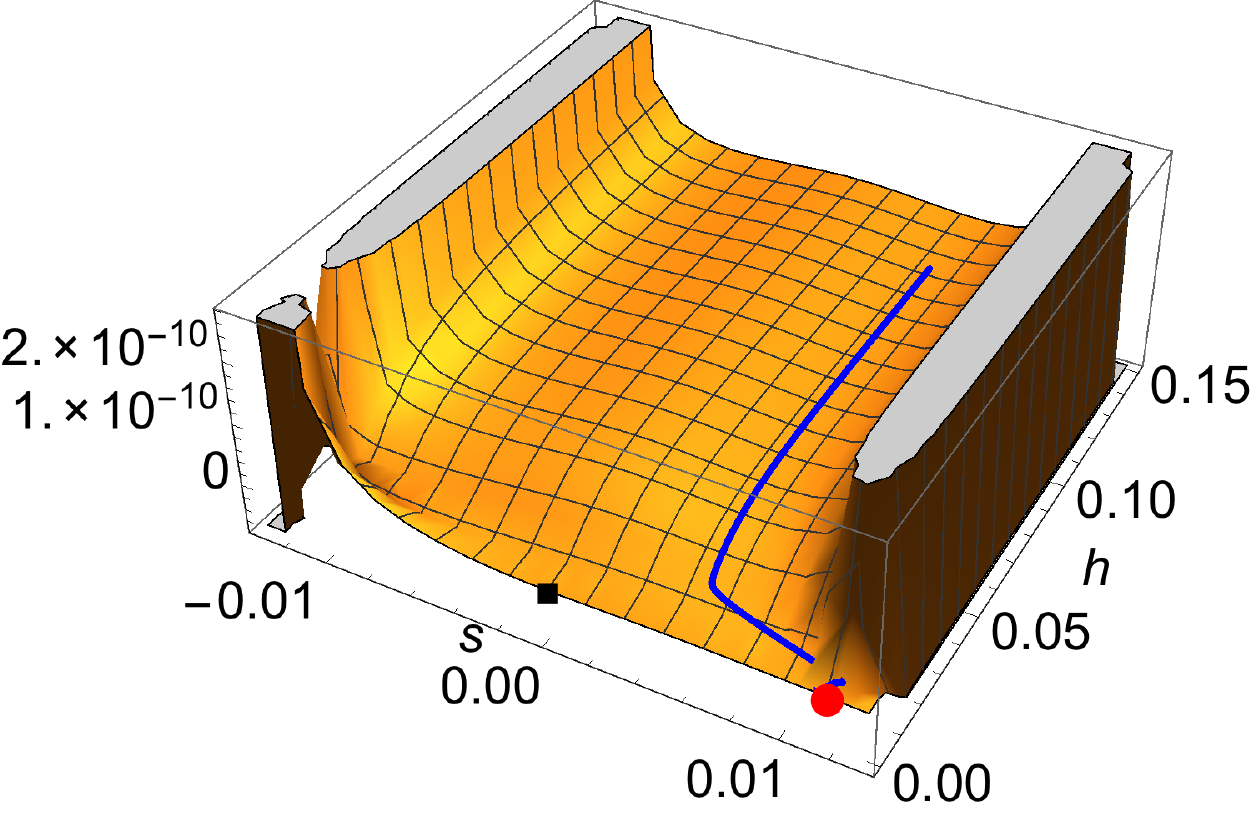}
\includegraphics[width=44mm]{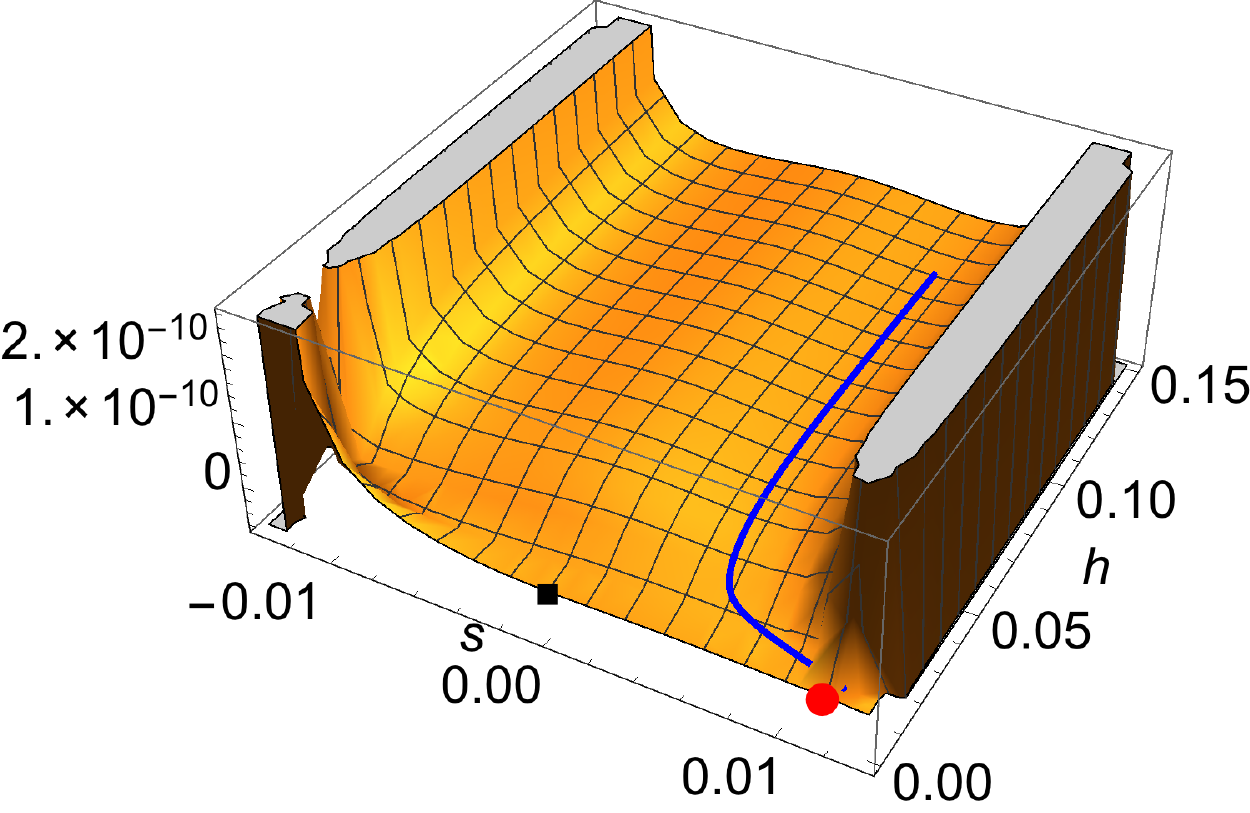}
\caption{\label{fig:sextic_traj}
Four types of inflaton trajectories:
(a) escape through the left wall, (b) settle at the GUT vacuum, (c) settle at the SM vacuum, (d) escape through the right wall.
The $\zeta$ parameter for (a), (b), (c) and (d) are respectively
$\zeta=-2800$, $-2964$, $-3050$ and $-3400$.
The parameter $\omega$ is $\omega=10^7$ for all cases.
}
\end{figure*}

\begin{figure*}[t]
\includegraphics[width=80mm]{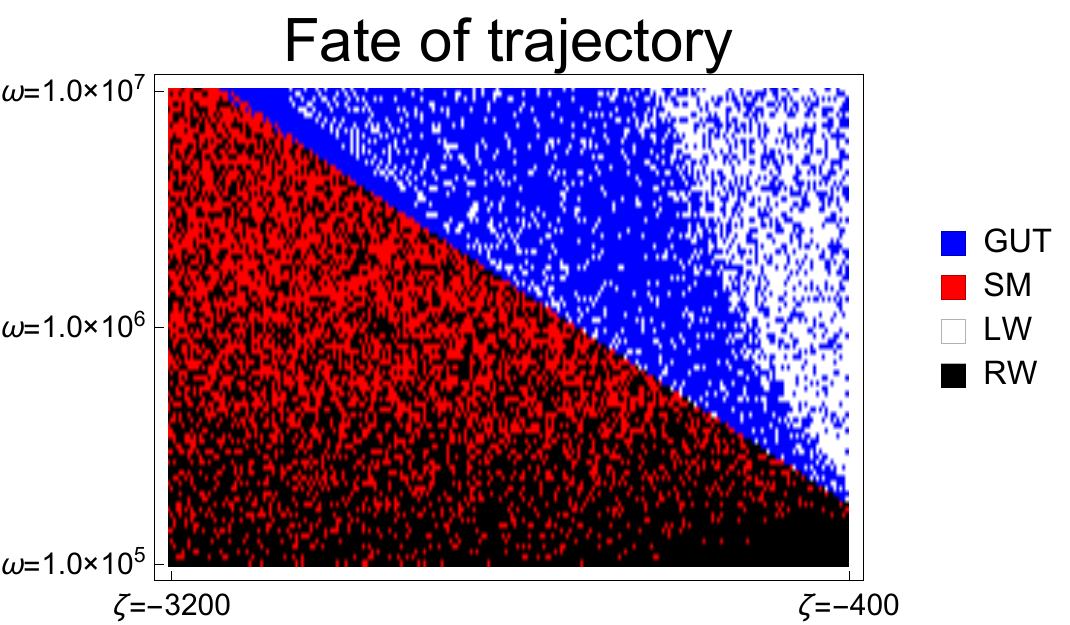}
\includegraphics[width=80mm]{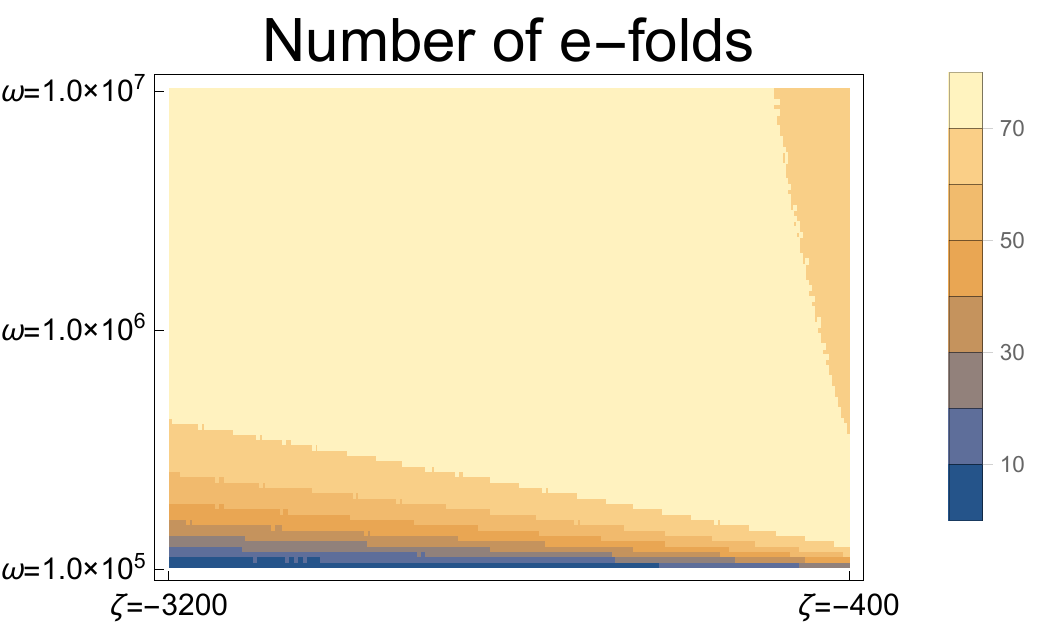}
\caption{\label{fig:sextic_solutions}
The four types of numerical solutions with different fates (left) and the
e-folding number (right) for numerical solutions in the parameter range
$-3200\leq\zeta\leq -400$ and $10^{5}\leq\omega\leq 10^{7}$.
These parameters are changed with the stepsize $\Delta\zeta=12$, $\Delta\omega=1.1\times 10^{5}$.
}
\end{figure*}

\subsubsection{Cosmological observables}
\label{subsubsec:cubic_CO}

Let us now discuss the inflationary predictions for this model.
For the reasons mentioned above, we focus on the benchmark inflaton trajectories that
end up in the SM vacuum and lie close to the separatrix line \eqref{eqn:CubicSeparatrix} in the parameter space $(\zeta,\omega)$.
Concretely, we choose $\omega=-0.0115 \, \zeta - 2.00$, a line slightly below the separatrix line \eqref{eqn:CubicSeparatrix}~\footnote{
In our grid search, the parameter values along the separatrix line \eqref{eqn:CubicSeparatrix} do not always give rise to the SM-type solutions.
}, and $2000\leq\zeta\leq 10000$.
Table \ref{tab:cubic} shows the field value $h=h_*$ at the horizon crossing for the e-folding $N_e=60$, $h=h_{\rm end}$ at the end of slow roll, the e-folding number $N_e$ between $h=h_{\rm init}=0.12$ and $h=h_{\rm end}$, the scalar and the tensor power spectra $\mathcal{P}_{\mathcal{R}}$ and $\mathcal{P}_{T}$, the scalar spectral index  $n_{s}$, the tensor-to-scalar ratio $r$ and the local-type nonlinearity parameter $f_{\rm NL}^{{\rm local}}$ for $\zeta=2000$, 4000, 6000, 8000, 10000.
Fig. \ref{fig:cubic_CO} shows the scalar power spectrum ${\C P}_{\C R}$, the scalar spectral index $n_s$, and the tensor-to-scalar ratio $r$ plotted against the parameter $\zeta$.
To obtain these results, we have once again solved the background equation forward in time from the initial values, identified the end point of inflation at which $\epsilon=1$, solved the background equation backward in time from the end point of inflation to find the horizon crossing with the $N_{e}=60$ condition, and then computed the cosmological observables.
See Appendix \ref{subsec:Fluc} and \ref{subsec:CosmoObs} for the technicalities and relevant formulas.
We have also computed the power spectrum for the isocurvature mode $\mathcal{P}_{\mathcal{S}}$, which is not shown in the table as it is found to be exponentially suppressed.
This suppression is due to the relatively large mass for the $s$ field introduced by the K\"ahler metric; it is found impossible to obtain a sensible inflaton trajectory without introducing a large mass for $s$ in this model.
The isocurvature fraction $\beta_{{\rm iso}}$ is thus essentially zero for these parameter values, which is 
consistent with observations \cite{Ade:2015xua}.

As the parameter $\zeta$ is varied, the scalar power spectrum $\mathcal{P}_{\mathcal{R}}$ changes somewhat; 
in view of the Planck 2015 (TT+TE+EE+lowP) results \cite{Ade:2015xua,Ade:2015lrj}
\begin{align}\label{eqn:PlanckPowerSpec}
2.133 \times 10^{-9}
\leq \mathcal{P}_{\mathcal{R}} \leq
2.283 \times 10^{-9}\,\quad
(68\% \,\, {\rm{C.L.}}),
\end{align}
only the range $\zeta \gtrsim 8356.9$ is observationally consistent but this certainly does not mean that lower values of $\zeta$ are not allowed in this model, as we have fixed the $\xi$ parameter using the Planck normalization of ${\C P}_{\C R}$ at $(\zeta,\omega)=(10000, -116)$.
The spectral index $n_{s}$ and the tensor-to-scalar ratio $r$ are, in contrast, found to be extremely insensitive to the change of $\zeta$.
These values of $n_{s}$ and $r$ are well inside the Planck constraints 
\cite{Ade:2015xua,Ade:2015lrj},
\begin{align}
	n_{s} = 0.9652 \pm 0.0047 \qquad
	(68\% \,\, {\rm C.L.})\,,
\end{align}
as well as the BICEP2/Keck Array/Planck results \eqref{eqn:rBKP}.
The nonlinearity parameter $f_{{\rm NL}}^{{\rm local}}$ (the local-type non-Gaussianity; see \cite{Kawai:2014gqa} for the detail of our numerical code based on the backward $\delta N$ formalism) is found to be $f_{{\rm NL}}^{{\rm local}}\sim\mathcal{O}(1)$, in the parameter region of interest.
For $\zeta\lesssim 6.3\times 10^3$, the nonlinearity parameter is outside the Planck constraints \cite{Ade:2015ava}~\footnote{
These constraints are from the temperature data alone.
},
\begin{align}
	f_{{\rm NL}}^{{\rm local}} = 2.5 \pm 5.7
	\qquad
	(68\% \,\, {\rm C.L.})\,,
	\label{eqn:NGPlanck2015}
\end{align}
but again this does not imply that the lower values of $\zeta$ are excluded as the $\xi$ parameter may be readjusted.

\subsection{The sextic K\"{a}hler model}

Let us next consider the other case in which the K\"{a}hler potential includes the noncanonical sextic term.
The nontrivial component of the K\"{a}hler metric is \eqref{eqn:kappasextic}, which vanishes when
$s^2=(-\zeta\pm\sqrt{\zeta^2+9\omega})/9\omega$.
Thus $\kappa=0$ at four values of $s$'s which are
(i) 
neither real nor pure imaginary when $\omega<-\zeta^2/9$;
(ii.a) all pure imaginary when $-\zeta^2/9<\omega<0$ and $\zeta<0$;
(ii.b) all real when $-\zeta^2/9<\omega<0$ and $\zeta>0$;
(iii) 2 real and 2 pure imaginary when $\omega>0$.
To remove the tachyonic instabilities in the $s$ direction, we must have 
$\kappa=0$ at at least two real $s$'s (these are the locations of the singularity walls).
Thus the parameter regions that are of our interest are
(ii.b) $-\zeta^2/9<\omega<0$, $\zeta>0$ and (iii) $\omega>0$.
We focus on the (iii) case below.

\begin{table*}[t]
\begin{center}\begin{tabular}{cc||cccccccc}
$\omega$ & $\zeta$ & $h_*$ & $h_{{\rm end}}$ & $N_e$ & ${\C P}_{\C R}$ & ${\C P}_T$ & $n_s$ & $r$ & $f_{\rm NL}^{{\rm local}}$\\ \hline
1.0$\times 10^{7}$ & -3068.6 & 0.1047 & 0.0080 & 79.67 & 2.185$\times 10^{-9}$ & 6.727$\times 10^{-12}$ & 0.9670 & 0.0031 & -1.046 \\
8.5$\times 10^{6}$ & -2625.2 & 0.1047 & 0.0081 & 79.66 & 2.214$\times 10^{-9}$ & 6.816$\times 10^{-12}$ & 0.9670 & 0.0031 & -1.124 \\
6.5$\times 10^{6}$ & -2034.1 & 0.1047 & 0.0083 & 79.65 & 2.255$\times 10^{-9}$ & 6.940$\times 10^{-12}$ & 0.9670 & 0.0031 & -1.230 \\
4.0$\times 10^{6}$ & -1442.9 & 0.1048 & 0.0087 & 79.60 & 2.300$\times 10^{-9}$ & 7.067$\times 10^{-12}$ & 0.9671 & 0.0031 & -1.607 \\
2.5$\times 10^{6}$ & -851.74 & 0.1049 & 0.0099 & 79.40 & 2.354$\times 10^{-9}$ & 7.199$\times 10^{-12}$ & 0.9671 & 0.0031 & -2.153 \\
5.0$\times 10^{5}$ & -260.58 & 0.1078 & 0.0158 & 75.61 & 2.568$\times 10^{-9}$ & 7.327$\times 10^{-12}$ & 0.9681 & 0.0029 & -3.295
\end{tabular}
\caption{
The values of the field $h$ at the horizon crossing $h_{*}$ and at the end of slow roll $h_{\rm end}$, the e-folding number $N_e$, the scalar and tensor power spectra ${\C P}_{\C R}$ and ${\C P}_T$, the scalar spectral index $n_s$, the tensor-to-scalar ratio $r$ and the local-type nonlinearity parameter $f_{\rm NL}^{{\rm local}}$ in the sextic K\"{a}hler model as the parameters $\zeta$ and $\omega$ are varied.
The initial value of the $h$ field is chosen to be $h_{\rm init}=0.12$ and the parameter $\xi$ is fixed to 6450 using the Planck normalization of the scalar power spectrum when $(\omega,\zeta)=(10^{7},-3000)$ and e-foldings 60.
The $N_e$ in the table is the e-folding number between $h_{\rm init}$ and $h_{{\rm end}}$.
}
\end{center}
\label{tab:sextic}
\end{table*}

\begin{figure*}[t]
\includegraphics[width=59mm]{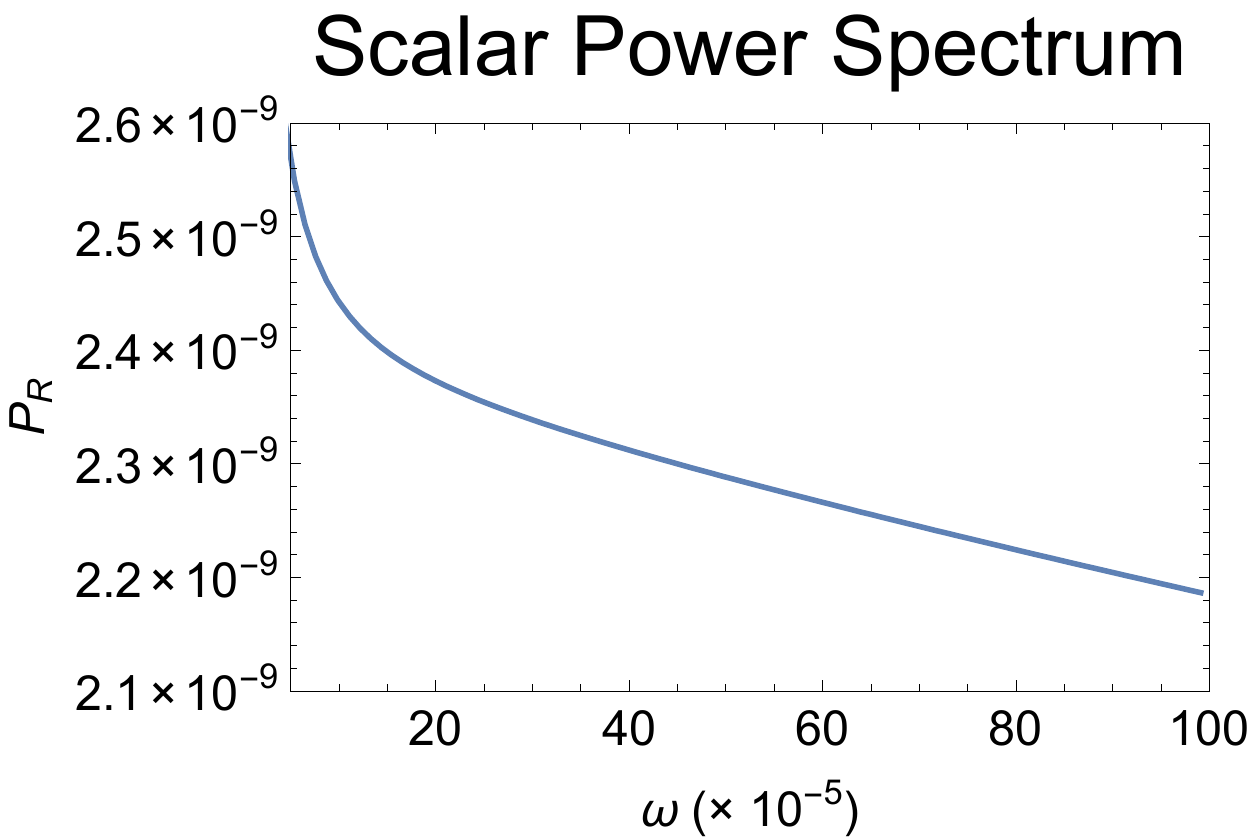}
\includegraphics[width=59mm]{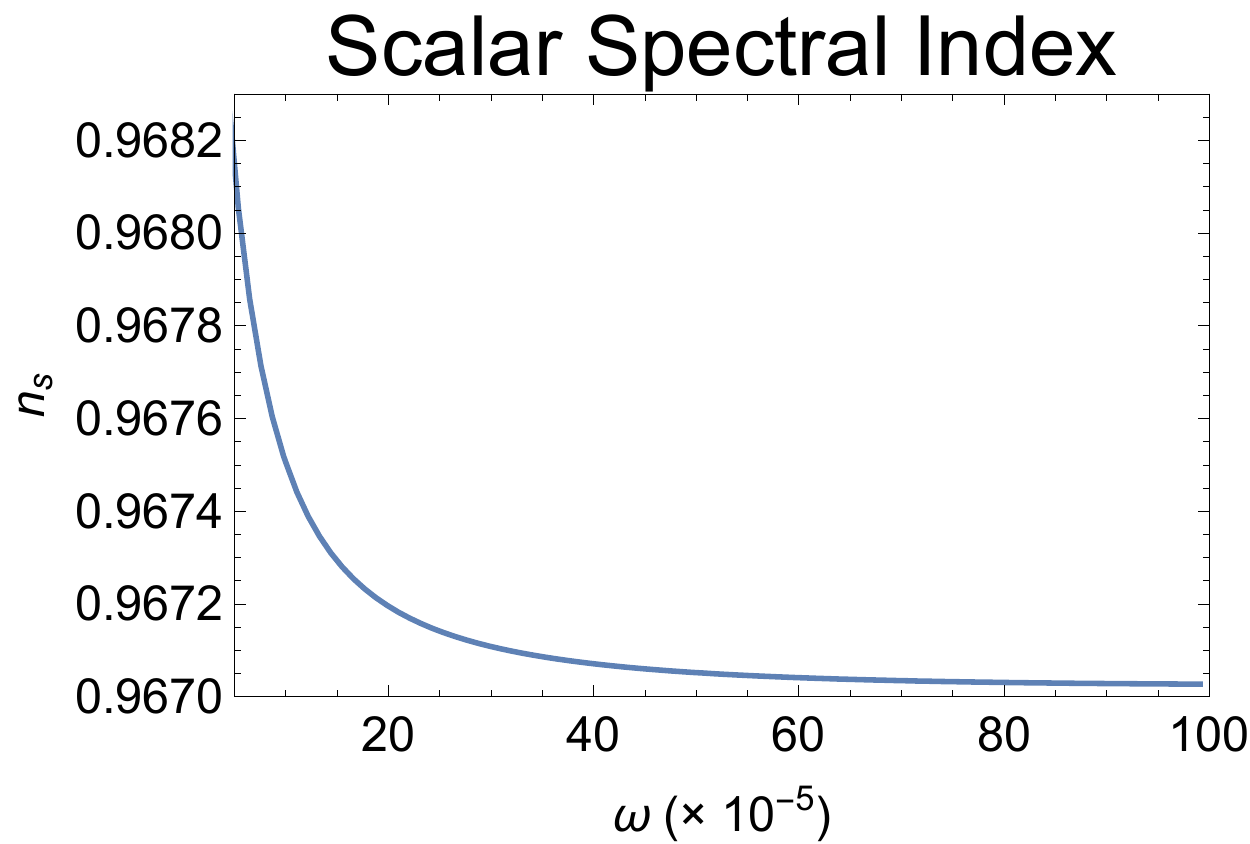}
\includegraphics[width=59mm]{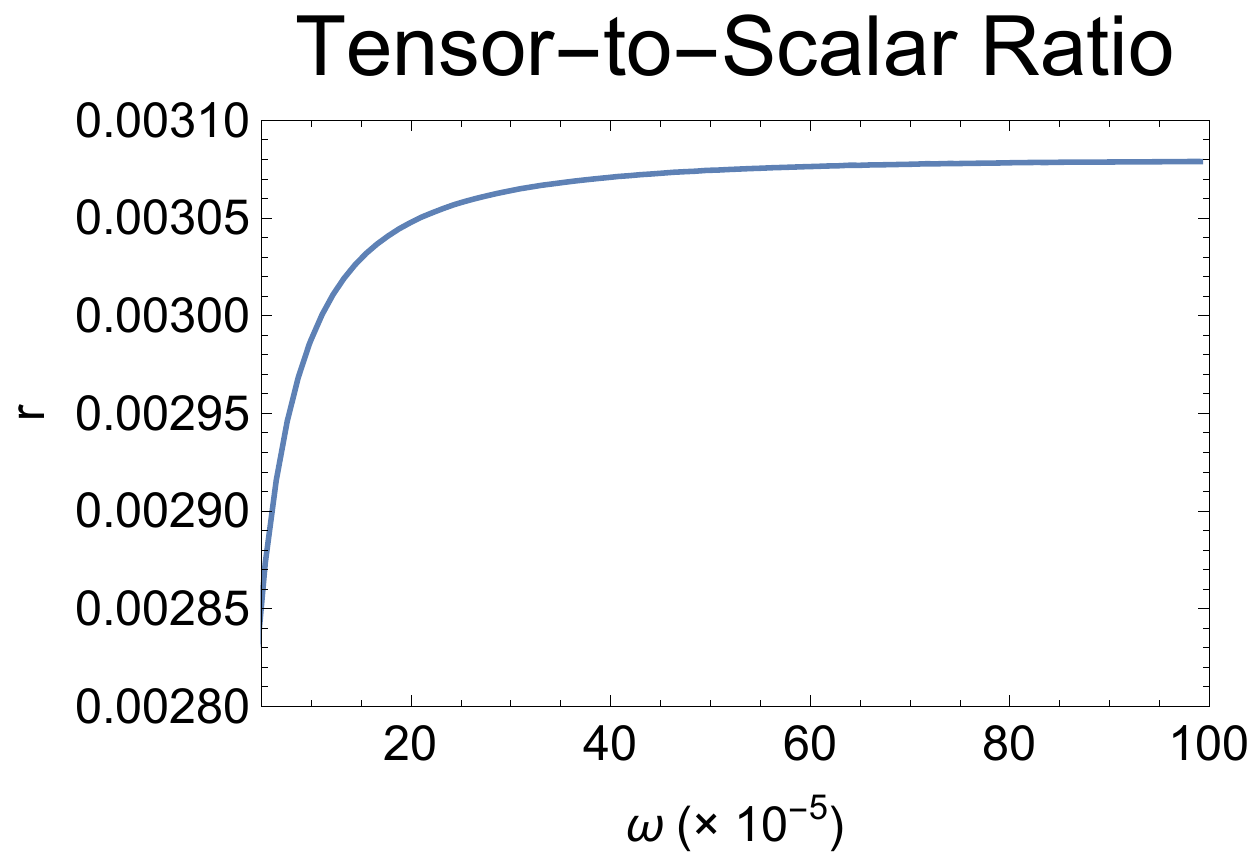}
\caption{\label{fig:sextic_CO}
Scalar power spectrum, scalar spectral index and tensor-to-scalar ratio for the sextic K\"{a}hler model.
}
\end{figure*}

\subsubsection{Background solutions}

Fig.~\ref{fig:sextic_stability} shows the behavior of the potential as the parameter $\omega$ is varied, when $\zeta$ is fixed to zero. 
Similarly to the cubic K\"ahler case, we choose the reference point $\zeta = -3000$ and 
$\omega = 10^{7}$ in the parameter space, for which the Planck normalization of the scalar power spectrum gives $\xi = 6450$.
We use this value of $\xi$ throughout the study of the sextic K\"{a}hler potential model.
$\omega=10^7$ is large enough to remove the tachyonic instabilities of the potential in the $s$ field direction.
For the trajectories shown in Fig.~\ref{fig:sextic_traj}, the initial value of $h$ is chosen to be the same as in the cubic case, $h=h_{{\rm init}} = 0.12$.

To study the behavior of the background solutions near the $(\zeta,\omega)=(-3000,10^{7})$ solution,
we solved, like in the cubic K\"ahler potential case, the background equations of motion fully numerically without slow-roll approximation.
We found the four types of inflaton trajectories LW, GUT, SM and RW similar to the cubic case.
Examples of these are shown in Fig.~\ref{fig:sextic_traj}.
The distribution of these four types numerical solutions in the parameter range of $-3200\leq\zeta\leq -400$ and $10^{5}\leq\omega\leq 10^{7}$ are shown in the left panel of Fig.~\ref{fig:sextic_solutions}.
Clear separation between the LW+GUT solutions and the SM+RW solutions can be seen; the
separatrix line is found to be
\begin{align}\label{eqn:SexticSeparatrix}
	\omega \approx -3.38 \times 10^{3} \zeta - 4.33 \times 10^{4}.
\end{align}
The right panel of Fig.~\ref{fig:sextic_solutions} shows the the number of e-folds between 
$h_{\rm init}=0.12$ and $h_{\rm end}$ at which the slow roll parameter $\epsilon$ becomes unity. 
Phenomenologically viable trajectories are those reaching the SM vacuum after the slow roll; 
they are below the separatrix \eqref{eqn:SexticSeparatrix}.
As emphasized in \ref{subsubsec:cubic_BG}, the escaping behavior of the LW and RW solutions are numerical artifacts and they should be considered as the GUT-type and the SM-type solutions, respectively.

\subsubsection{Cosmological observables}

To compute the cosmological observables in this model, we adopt the same methodology as explained in \ref{subsubsec:cubic_CO}.
We focus on the parameter region near the separatrix \eqref{eqn:SexticSeparatrix} and change the value of $\omega$.
Concretely, we choose 
$\zeta = -2.96 \times 10^{-4}\, \omega - 112.8$ and
$4.3\times 10^{5} \leq \omega \leq 1.0\times 10^{7}$.
This line is slightly below the separatrix line \eqref{eqn:SexticSeparatrix}, as 
the parameters exactly on the separatrix do not always give the SM-type solutions in the grid search;
we are only interested in the SM-type trajectories that are phenomenologically viable. 
The lower value of $\omega=4.3 \times 10^{5}$ is chosen to yield at least 60 e-foldings with our initial conditions $h_{\rm init}=0.12$ (see Fig.~\ref{fig:sextic_solutions}).

In Fig.~\ref{fig:sextic_CO} the scalar power spectrum $\mathcal{P}_{\mathcal{R}}$, its spectral index $n_{s}$ and the tensor-to-scalar ratio $r$ are plotted for different values of $\omega$ ($\zeta$ is chosen to be on the line above and $\xi$ is fixed).
We again found that the isocurvature fraction $\beta_{{\rm iso}}$ is negligible, for the same reason as in the cubic K\"ahler potential case.
Table~\ref{fig:sextic_CO} shows the field value of $h$ at the horizon crossing, the value of $h$ at the end of inflation, the e-folding number $N_{e}$ between $h=h_{\rm init}$ and at the end of the slow roll, the power spectra $\mathcal{P}_{\mathcal{R}}$ and $\mathcal{P}_{T}$, the scalar spectral index $n_{s}$, the tensor-to-scalar ratio $r$ and the local-type nonlinearity parameter $f_{{\rm NL}}^{{\rm local}}$ for several sample values of $(\zeta,\omega)$.
The values of $n_{s}$ and $r$ in the table are well within the constraints of the latest Planck experiment results \cite{Ade:2015lrj} we well as the BICEP2/Keck Array/Planck joint results \cite{Ade:2015xua}.
For lower values of $\omega$ the scalar power spectrum is seen to increase and goes outside the Planck constraints \eqref{eqn:PlanckPowerSpec} for $\omega \lesssim 52.4 \times 10^{5}$, but this is not meant to be the lower bound of this parameter as the $\xi$ parameter may be readjusted 
(recall that we have fixed $\xi=6450$ at $(\zeta,\omega)=(-3000,10^{7})$ using the Planck normalization).
The local-type nonlinearity parameter is $f_{{\rm NL}}^{{\rm local}} \sim \mathcal{O}(1)$ throughout the parameter region of interest and is marginally within the present observatoinal constraints \eqref{eqn:NGPlanck2015}.

\subsection{Summary}

We have seen in this section the behavior of the background inflaton trajectories and the prediction for the cosmological observables in the inflationary scenarios introduced in the previous section.
For systematic parameter scan, we fixed the $\xi$ parameter by the Planck normalization of the scalar power spectrum at special points in the parameter spaces:
$(\zeta,\omega)=(10000,-116)$ for the cubic K\"ahler case and $(\zeta,\omega)=(-3000, 10^7)$ in the sextic K\"ahler case.
We then varied the parameters $\zeta$ and $\omega$, in the range of
$2000\leq\zeta\leq10000$, $-200\leq\omega\leq 0$ for the cubic K\"ahler model, and
$-3200\leq\zeta\leq-400$, $10^{5}\leq\omega\leq10^{7}$ for the sextic K\"ahler model.
In both cases we obtained four types of numerical solutions: 
two types of runaway solutions LW and RW, and the one that ends up in the GUT vacuum and the other that ends up in the SM vacuum.
The runaway solutions are due to the (unavoidable) pathological behavior of the numerical integration near the K\"ahler metric singularities; these walls, while infinitely high, become infinitesimally thin near the measure-zero pin holes.
Since a classical trajectory cannot penetrate such a wall, these runaway solutions should be regarded as numerical artifacts.

The observable parameters predicted in the cubic and the sextic K\"ahler models are quite similar. We have selected the phenomenologically viable and numerically well-behaved sets of background inflaton trajectories on the SM side of the separatrix between the GUT- and SM-type solutions in the $\zeta$-$\omega$ plane.
We then computed the scalar power spectra in the adiabatic and the isocurvature modes, the tensor power spectrum, the scalar spectral index, the tensor-to-scalar ratio, and the local-type nonlinearity parameter for these inflaton trajectories.
A well-known attractive feature of the Bezrukov-Shaposhnikov scenario of Higgs inflaiton is that the prediction for the scalar spectral index and the tensor-to-scalar ratio agrees remarkably well with the present observations \cite{Ade:2015xua,Ade:2015lrj}, once the nonminimal coupling parameter $\xi$ is fixed by the scalar power spectrum.
This feature is found to persist in our supersymmetric GUT embedding, in both cubic and sextic K\"ahler potential cases.
Supersymmetric GUT embedding necessarily involves multiple scalars and in principle the multifield effects may change the cosmological observables; we have found that such effects, in particular the isocurvature mode of fluctuations, are negligible in our model.
The absence of the isocurvature mode is due to the large effective mass along the $s$ field direction.
We also found that the nonlinearity parameter is ${\C O}(1)$.
As we vary the parameters $\zeta$ and $\omega$ along the vicinity of the separatrix, the scalar power spectrum ${\C P}_{\C R}$ is found to deviate from the Planck-normalized value, while the spectral index $n_{s}$ and the tensor-to-scalar ratio $r$ are insensitive to the change of these parameters.
The isocurvature fraction $\beta_{{\rm iso}}$ stays negligible and $f_{{\rm NL}}^{{\rm local}}$ stays ${\C O}(1)$, as long as the scalar power spectrum stays close to the Planck-normalized value.

\section{Discussion}\label{sec:concl}

As a well-motivated and technically natural beyond-the-SM implementation of the Bezrukov-Shaposhnikov scenario, we have discussed, extending the work of \cite{Arai:2011nq}, Higgs inflation in supersymmetric GUT in this paper. 
The supergravity $\eta$ problem is avoided by using the noncanonical K\"ahler potential in the superconformal framework; the noncanonical term gives rise to the nonminimal coupling of the Higgs field as in the Bezrukov-Shaposhnikov scenario.
We have considered the minimal $SU(5)$ GUT model as a prototype of GUT and analyzed the model including multifield effects. 
The prediction for the scalar spectral index $n_s$ and the tensor-to-scalar ratio $r$ is very similar to the Bezrukov-Shaposhnikov scenario of SM Higgs inflation and thus agrees very well with the present observations. 
This feature is found to be insensitive to the change of the model parameters 
$\zeta$ and $\omega$, which may indicate some attractor mechanism, similar to the one studied recently in \cite{Kallosh:2013maa,Kallosh:2013yoa,Kallosh:2013daa}.
The prediction of cosmological parameters in this model is also robust against multifield ambiguities, as the isocurvature mode of the fluctuations is found to be negligible.

In the supersymmetric $SU(5)$ Higgs inflation model that we have studied, the non-Gaussianity (the local-type nonlinearity parameter) stays relatively small, reflecting the fact that the multifield effects are overall insignificant.
In similar embedding of Higgs inflation in the NMSSM or in the supersymmetric seesaw, in contrast, the non-Gaussianity can be important \cite{Kawai:2014gqa}.
Why are the effects less important in the $SU(5)$ case?
A salient feature of inflation models realized in grand unification is that the GUT symmetry is broken during inflation.
In the scenario we have studied, this is related to the asymmetry of the inflaton potential in the singlet (the $s$ field) direction; the requirement that the trajectory must reach the SM vacuum disfavors trajectories that typically produce large isocurvature modes and large non-Gaussianity, that is, trajectories that stay on a ridge of the potential for a while and then make a turn \cite{Peterson:2010np,Peterson:2010mv}.
While symmetries of an inflaton potential is commonly imposed in simple toy examples of multifield inflationary models, one cannot expect high symmetries in generic inflationary models, such as in GUT scenarios or the stringy landscape scenario (see however \cite{Kofman:2004yc} for a discussion in favor of symmetries in generic models).
We have provided a concrete case study of a GUT scenario in this paper and found that multifield effects are not important.
The results seem to indicate that discussions based on inflationary toy models tend to overestimate the multifield effects.

Let us conclude by commenting on possible directions of further research.
One direction is to investigate less trivial examples of GUT embedding.
While the $SU(5)$ GUT is widely recognized as a prototype of grand unification, it is certainly not an entirely satisfactory example as it suffers e.g. from the proton decay problem.
While many of the features found here are expected also in other GUT models, 
quantitative consistency check on various phenomenological and cosmological bounds in concrete realistic scenarios is certainly desirable.
Another important topic that we have not touched upon in this paper is (p)reheating after inflation\footnote{The reheating process of the SM Higgs inflation is discussed in \cite{GarciaBellido:2008ab}.}.
Recent studies of nonminimally coupled multifield reheating based on simple inflationary toy models indicate that energy transfer due to parametric resonance 
is efficient \cite{DeCross:2015uza}, since the strong single-field attractor behavior persists during reheating and multifield de-phasing effects can 
be avoided.
In our phenomenological example based on supersymmetric GUT, in contrast, the inflaton dynamics after the slow roll may exhibit irregular and chaotic behavior, due to the irregular shape of the scalar potential near the GUT and the SM vacua.
Such irregular motion may lead to suppression of the resonance effects as in \cite{Battefeld:2008bu}.

\acknowledgments

We acknowledge helpful conversations with Nobuchika Okada.
We also thank the authors of TransportMethod \cite{Dias:2015rca} and 
MultiModeCode \cite{Price:2014xpa} for helpful communications.
This work was supported in part by the National Research Foundation of Korea Grant-in-Aid for Scientific Research No. NRF-2015R1D1A1A01061507 (S.K.) and the NRF Global Ph.D. Fellowship Program No. 2011-0008792 (J.K.). We used computing resources of the Yukawa Institute, Kyoto University.

\appendix

\section{Multi-field inflaton dynamics}\label{sec:dynamics}
%

To analyze the two-field inflation model given by the Lagrangian \eqref{eqn:LE}, we first describe the background solutions and then discuss fluctuations on the background.
We use the covariant formalism developed in \cite{Sasaki:1995aw,Nakamura:1996da,Gordon:2000hv,GrootNibbelink:2001qt,Peterson:2010np,Gong:2011uw,Kaiser:2012ak,Schutz:2013fua}.
We denote the background inflaton fields as $\varphi^a$
and the fluctuations as $\delta\phi^a$, i.e.
\begin{align}
\phi^a(t,x)=\varphi^a(t)+\delta\phi^a(t,x).
\end{align}
The background inflaton is assumed to have no spatial dependence.

\subsection{Background solutions}
\label{subsec:BkgSol}

The Klein-Gordon equation obtained from the Lagrangian \eqref{eqn:LE} is
\begin{align}\label{eqn:KG}
\frac{D\dot\varphi^a}{dt}+3H\dot\varphi^a+G^{ab}\nabla_b V(\varphi^c)=0,
\end{align}
where $\frac{D}{dt}$ is the covariantized time derivative which is defined to operate on a field space vector $X^a$ as
\begin{align}
\frac{DX^a}{dt}=\dot X^a+\Gamma^a{}_{bc} X^b\dot\varphi^c.
\end{align}
The overdot is the derivative with respect to the cosmic time $t$.
$H=\dot a/a$ is the Hubble parameter.
$\nabla_a$ denotes the covariant derivative in the field space; its connection is the Christoffel symbol
$\Gamma^a{}_{bc}$
(\eqref{eqn:CubicGamma} or \eqref{eqn:SexticGamma})
constructed from the field space metric (\eqref{eqn:FSmetric_cubic} or \eqref{eqn:FSmetric_sextic}).
The Einstein equations are
\begin{align}
&3H^2=\rho,\label{eqn:Friedmann}\\
&\dot\rho+3H(\rho+P)=0,\label{eqn:EnergyConserve}
\end{align}
where
\begin{align}
\rho\equiv\half G_{ab}\dot\varphi^a\dot\varphi^b+V(\varphi^c),\label{eqn:KG1}\\
P\equiv\half G_{ab}\dot\varphi^a\dot\varphi^b-V(\varphi^c),\label{eqn:KG2}
\end{align}
are the energy density and the effective pressure of the inflaton fields.
In numerics we solve the Klein-Gordon equation \eqref{eqn:KG} and use the Friedmann equation \eqref{eqn:Friedmann} to obtain the Hubble parameter.
The equation of energy conservation \eqref{eqn:EnergyConserve} is used to monitor numerical accuracy.

To solve the inflaton dynamics in the flat Friedmann Universe we need the initial conditions for
$\varphi^a=(s,h)$ and $\dot\varphi^a=(\dot s, \dot h)$.
In this paper we are considering the inflationary model driven by the SM Higgs field $h$.
Hence, the initial value of $h$ is assumed to be large, $h(t_0)\gg s(t_0)$.
Turning on the $\zeta$ parameter (the quartic term of the K\"{a}hler potential), the inflaton potential
becomes a half pipe shape along $h$, making Higgs-driven inflation possible \cite{Arai:2011nq}.
Such a half pipe however cannot be too narrow since the SM vacuum $s=v\sim M_{\rm GUT}$, $h=0$ needs to be reached from the initial point $s=s(t_0), h=h(t_0)$.
During inflation, light fields have quantum fluctuations of the order of the Hubble parameter.
For example, the $s$ field is expected to have fluctuations
\begin{align}
\langle(\Delta\hat s)^2\rangle\approx\langle G_{ss}(\Delta s)^2\rangle\approx \frac{H^2}{(2\pi)^2},
\end{align}
where $\hat s$ is canonically normalized in the Einstein frame.
Thus $|\Delta s|\approx\frac{H}{2\pi}\sqrt{1+\xi h}$, assuming $\langle s\rangle\ll M_{\rm P}$.
Fine-tuning the initial value of $s$ below this $\Delta s$ is considered unnatural.
In the numerical study we assume the inflaton already follows an attractor at 60 e-folds before the end of inflation; we thus choose $s(t_0)$ to be at a local minimum for a given (large enough) $h(t_0)$. 
We may consistently set $\dot s(t_0)=0$.
$\dot h(t_0)$ is found by the slow roll equation of motion.

\subsection{Fluctuations}
\label{subsec:Fluc}

It is well-known that the ten degrees of freedom (DoF) in the metric perturbation on the Friedmann-Lema\^{i}tre-Robertson-Walker (FLRW) background split into a tensor (2 DoF), two vectors (4 DoF), and four scalars (4 DoF) of $SO(3)$.
These include 2 gauge DoF in the vector and 2 gauge DoF in the scalar mode.
The 2 DoF in the tensor mode represent the two helicity states of gravitational waves.
After the horizon exit, the tensor mode fluctuations undergo no nontrivial evolution as they decouple from the scalar mode.
In the absence of a vector source, the evolution of the vector mode fluctuations is trivial decay \cite{Maleknejad:2012fw} and hence of no interest; we will not discuss them any further.
The metric with scalar mode fluctuations $A$, $B$, $E$, $\psi$ and tensor mode fluctuations $h_{ij}$ may be written as
\begin{align}
	ds^2=&-(1+2A)dt^2+2a B_{|i}dt dx^i\nn\\
	&+a^2\left[
	(1-2\psi)\delta_{ij}+2E_{|ij}+h_{ij}\right]
	dx^idx^j,
\end{align}
where $a=a(t)$ is the scale factor and ${*}_{|i}\equiv\frac{\partial *}{\partial x^i}$
is the spatial derivative with respect to the comoving coordinates.
In multifield inflation with $n$ ($=2$ in our case) inflaton components, there are $n+4$ scalar DoF, 2 of which are the gauge DoF.
Since there are two constraint equations, the physical scalar DoF is $n$.
This indicates that the dynamics of scalar mode fluctuations may be studied by analyzing essentially the perturbed Klein-Gordon equations for the inflaton fields alone.
The relevant equations of motion for the scalar perturbations are neatly expressed by using the covariant formalism \cite{Sasaki:1995aw,Nakamura:1996da,Gordon:2000hv,GrootNibbelink:2001qt,Peterson:2010np,Kaiser:2012ak,Schutz:2013fua} as
\begin{align}
	&\frac{D^2 Q^a}{dt^2}+3H\frac{DQ^a}{dt}+
	\nn\\
	&\qquad
	\left[
	\frac{k^{2}}{a^{2}}\delta^{a}{}_{b}
	+\mathcal{M}^{a}{}_{b}
	-\frac{1}{a^{3}}\frac{D}{dt}\left(
	\frac{a^{3}}{H}\dot{\varphi}^{a}\dot{\varphi}_{b}
	\right)
	\right]Q^{b}=0\,,
	\label{eqn:EOMfluc}
\end{align}
where
\begin{align}
	Q^{a} = \mathcal{Q}^{a}
	+\frac{\dot{\varphi}^{a}}{H}\psi\,,
\end{align}
is the Mukhanov-Sasaki gauge-invariant variable and $\mathcal{Q}^{a}$ is a covariant fluctuation vector which is related to $\delta\phi^{a}$ to first order in the fluctuations.
The effective mass-squared matrix $\mathcal{M}^{a}{}_{b}$ is defined as
\begin{align}\label{eqn:effMsq}
	\mathcal{M}^{a}{}_{b}
	\equiv
	G^{ac}\nabla_{b}\nabla_{c}V
	-R^{a}{}_{cdb}\dot{\varphi}^{c}\dot{\varphi}^{d}\,,
\end{align}
where $R^a{}_{bcd}$ is the Riemann curvature tensor of the field space (see Eqs.~\eqref{eqn:Ricciscalar_cubic}, \eqref{eqn:Ricciscalar_sextic}).

One may decompose the fluctuations into the adiabatic and isocurvature components respectively as
\begin{align}
	Q_\parallel
	\equiv
	G_{ab}
	\sigma_{\parallel}^{a}
	Q^{b}
	\,,\quad
	Q_\perp
	\equiv
	G_{ab}
	\sigma_{\perp}^{a}
	Q^b\,,
\end{align}
where $\sigma_{\parallel}^{a}$ is the unit vector along the inflationary trajectory and $\sigma_{\perp}^{a}$ is the unit vector orthogonal to it.
These unit vectors are defined as
\begin{align}
	\sigma_{\parallel}^{a} &\equiv
	\frac{\dot{\varphi}^{a}}{|\dot{\varphi}|}
	=
	\frac{\dot{\varphi}^{a}}
	{\sqrt{G_{cd}\dot{\varphi}^{c}\dot{\varphi}^{d}}}
	\,,\\
	\sigma_{\perp}^{a} &\equiv
	\frac{D\sigma_{\parallel}^{a}/dt}{|D\sigma_{\parallel}^{a}/dt|}\,.
\end{align}
The absolute value of a quantity, say $X^{a}$, should be understood as $|X| \equiv \sqrt{G_{ab}X^{a}X^{b}}$.
The evolution equations \eqref{eqn:EOMfluc} are then written as
\begin{align}
	&\ddot Q_\parallel+3H\dot Q_\parallel+
	\nn\\
	&\qquad
	\left[
	\frac{k^{2}}{a^{2}}
	+\mathcal{M}_{\parallel\parallel}
	-\bigg\vert
	\frac{D\sigma_{\parallel}^{a}}{dt}
	\bigg\vert^{2}
	-\frac{1}{a^{3}}\frac{d}{dt}\left(
	\frac{a^{3}|\dot{\varphi}|^{2}}{H}
	\right)
	\right]Q_{\parallel}
	\nn\\
	&\qquad
	=
	2\frac{d}{dt}\left(
	\bigg\vert
	\frac{D\sigma_{\parallel}^{a}}{dt}
	\bigg\vert Q_{\perp}
	\right)
	-2\left(
	\frac{\nabla_{\sigma_{\parallel}}V}{|\dot{\varphi}|}
	+\frac{\dot{H}}{H}
	\right)
	\bigg\vert
	\frac{D\sigma_{\parallel}^{a}}{dt}
	\bigg\vert
	Q_{\perp}\,,
	\label{eqn:pertEOM1}
	\\
	&\ddot Q_\perp+3H\dot Q_\perp+
	\nn\\
	&\qquad
	\left[
	\frac{k^{2}}{a^{2}}
	+\mathcal{M}_{\perp\perp}
	+3\bigg\vert
	\frac{D\sigma_{\parallel}^{a}}{dt}
	\bigg\vert^{2}
	\right]Q_{\perp}
	=4\frac{1}{|\dot{\varphi}|}
	\bigg\vert
	\frac{D\sigma_{\parallel}^{a}}{dt}
	\bigg\vert
	\frac{k^{2}}{a^{2}}\Psi\,,
	\label{eqn:pertEOM2}
\end{align}
where
\begin{align}
	\mathcal{M}_{\parallel\parallel}
	&\equiv
	G_{ab}\sigma_{\parallel}^{b}
	\sigma_{\parallel}^{c}
	\mathcal{M}^{a}{}_{c}\,,\\
	\mathcal{M}_{\perp\perp}
	&\equiv
	G_{ab}\sigma_{\perp}^{b}
	\sigma_{\perp}^{c}
	\mathcal{M}^{a}{}_{c}\,,
\end{align}
and $\Psi$ is the gauge-invariant Bardeen potential,
\begin{align}
	\Psi \equiv
	\psi + a^{2}H\left(
	\dot{E}-\frac{B}{a}
	\right)\,.
\end{align}
We see from Eqs.~\eqref{eqn:pertEOM1} and \eqref{eqn:pertEOM2} that the adiabatic mode may be sourced by the isocurvature mode but
not vice versa.

In terms of $Q_{\parallel}$ and $Q_{\perp}$, the curvature perturbation and the isocurvature perturbation are given by
\begin{align}
	{\C R}&=\frac{H}{|\dot{\varphi}|}Q_\parallel,\\
	{\C S}&=\frac{H}{|\dot{\varphi}|}Q_\perp\,.
\end{align}
To study the evolution of fluctuations on superhorizon scales, it proves useful to introduce so-called transfer functions, defined as
\begin{align}
	T_{\mathcal{R}\mathcal{S}}(t_{*},t)
	&=
	\int_{t_{*}}^{t} dt^{\prime}\,
	\alpha(t^{\prime})
	H(t^{\prime})
	T_{\mathcal{S}\mathcal{S}}(t_{*},t^{\prime})\,,\\
	T_{\mathcal{S}\mathcal{S}}(t_{*},t)
	&=
	\exp\left[
	\int_{t_{*}}^{t} dt^{\prime}\,
	\beta(t^{\prime})H(t^{\prime})
	\right]\,,
\end{align}
where
\begin{align}
	\alpha &\equiv \frac{2}{H}
	\Big\vert
	\frac{D}{dt}\left(
	\frac{\dot{\varphi}^{a}}{
	|\dot{\varphi}|}
	\right)
	\Big\vert\,,\\
	\beta &\equiv
	-2\epsilon-\eta_{\perp\perp}+\eta_{\parallel\parallel}-\frac{4}{3H^{2}}
	\Big\vert
	\frac{D}{dt}\left(
	\frac{\dot{\varphi}^{a}}{
	|\dot{\varphi}|}
	\right)
	\Big\vert^{2}\,,
\end{align}
and $t_{*}$ is the horizon-crossing time at which $k=aH$.
We have introduced the slow-roll parameters $\epsilon$, $\eta_{\perp\perp}$ and $\eta_{\parallel\parallel}$ defined by
\begin{align}\label{eqn:epsilon}
	\epsilon &\equiv
	-\frac{\dot{H}}{H^{2}}\,,\\
	\eta_{\perp\perp} &\equiv
	\frac{\mathcal{M}_{\perp\perp}}{V}\,,\\
	\eta_{\parallel\parallel} &\equiv
	\frac{\mathcal{M}_{\parallel\parallel}}{V}\,.
\end{align}
The curvature and isocurvature perturbations at time $t$ after the horizon exit are then given by
\begin{align}\left(\begin{array}{c}
	{\C R}(t)\\
	{\C S}(t)
	\end{array}\right)
	=\left(\begin{array}{cc}
	1& T_{{\C R}{\C S}}(t_*,t)\\
	0& T_{{\C S}{\C S}}(t_*,t)
	\end{array}\right)
	\left(\begin{array}{c}
	{\C R}(t_*)\\
	{\C S}(t_*)
	\end{array}\right).
\end{align}

For the appropriateness of the approximations used see e.g. \cite{Gordon:2000hv,GrootNibbelink:2001qt,Peterson:2010np,Kaiser:2012ak,Schutz:2013fua}.

\subsection{Cosmological observables}
\label{subsec:CosmoObs}
The power spectra of curvature and isocurvature perturbations after the horizon exit are given in terms of the transfer functions as
\begin{align}
	{\C P}_{\C R}(k)=&{\C P}_{\C R}(k_*)\left[1+T_{\C R\C S}^2(t_*,t)\right],\\
	{\C P}_{\C S}(k)=&{\C P}_{\C S}(k_*)T_{\C S\C S}^2(t_*,t).
\end{align}
The curvature and isocurvature power spectra at horizon-crossing are
\begin{align}
	\mathcal{P}_{\mathcal{R}}(k_{*})
	=\mathcal{P}_{\mathcal{S}}(k_{*})
	=\left(
	\frac{H_*}{2\pi}
	\right)^{2}\frac{1}{2\epsilon_*}\,,
\end{align}
where $H_*$ and $\epsilon_*$ are evaluated at $t_{*}$.
The isocurvature fraction $\beta_{{\rm iso}}$ then becomes
\begin{align}
	\beta_{\rm iso}\equiv\frac{\C P_{\C S}}{\C P_{\C S}+\C P_{\C R}}
	=\frac{T_{\C S\C S}^2}{T_{\C S\C S}^2+T_{\C R\C S}^2+1}\,.
\end{align}

The tensor power spectrum is
\begin{align}
	\mathcal{P}_{T}=
	8\left(
	\frac{H_{*}}{2\pi}
	\right)^{2}\,.
\end{align}
It will not evolve in the superhorizon scales.

We use the standard definition of the scalar spectral index,
\begin{align}
	n_{s} \equiv 1+\frac{d\ln\mathcal{P}_{\mathcal{R}}}{d\ln k}\,.
\end{align}
It is evaluated as
\begin{align}
	n_{s} = n_{s,*}
	-\big(
	\alpha_{*}
	+\beta_{*}T_{\mathcal{R}\mathcal{S}}
	\big)\sin(2\Delta)\,,
\end{align}
where
\begin{align}
	n_{s,*} = 1-6\epsilon+2\eta_{\parallel\parallel}
\end{align}
is the spectral index at the horizon crossing and the angle $\Delta$ is defined by
\begin{align}
	\cos\Delta \equiv
	\frac{T_{\mathcal{R}\mathcal{S}}}{\sqrt{1+T_{\mathcal{R}\mathcal{S}}^{2}}}\,.
\end{align}
Finally the tensor-to-scalar ratio
\begin{align}
	r \equiv \frac{\mathcal{P}_{T}}{\mathcal{P}_{\mathcal{R}}}
\end{align}
is evaluated as
\begin{align}
	r = \frac{16\epsilon}{1+T_{\mathcal{R}\mathcal{S}}^{2}}\,.
\end{align}

%


\end{document}